\documentclass[]{aa}  

\usepackage{graphicx}
\usepackage{txfonts}
\usepackage{hyperref}
\usepackage{upgreek}
\usepackage{color}
\usepackage{comment}
\usepackage{mathtools}
\usepackage{ulem}
\usepackage{xspace}

\newcommand{\sectionname}{Sect.}
\newcommand{\dd}{\textrm{d}}

\newcommand{\deriv} [2] {\frac {\textrm{d} #1 } {\textrm{d} #2} }
\newcommand{\derivs} [2] {\frac {\textrm{d}^2 #1 } {\textrm{d} #2^2} }

\newcommand{\eqna} [1] {
\begin{eqnarray} #1
\end{eqnarray}}
\newcommand{\algn} [1] {
\begin{align} #1
\end{align}}

\let\originaleqref\eqref
\renewcommand{\eqref}{Eq.~\originaleqref}
\newcommand{\eqss}[2]{Eqs.~(\ref{#1})-(\ref{#2})}

\makeatletter
\newcommand{\eq}[1]{%
    \checknextarg{#1}}
\newcommand{\checknextarg}[1]{\@ifnextchar\bgroup{Eqs.~(\ref{#1})\gobblenextarga}{Eq.~(\ref{#1})\xspace}}
\newcommand{\gobblenextarga}[1]{\@ifnextchar\bgroup{, (\ref{#1})\gobblenextargb}{ and (\ref{#1})\xspace}}
\newcommand{\gobblenextargb}[1]{\@ifnextchar\bgroup{, (\ref{#1})\gobblenextargb}{, and (\ref{#1})\xspace}}
\makeatother

\begin{document} 

%%%%%%%%%%%%%%%%%%%

   \title{Probing the mid-layer structure of red giants}

   \subtitle{I. Mixed-mode coupling factor as a seismic diagnosis}
  \titlerunning{Mid-layer structure of red giants I} 
  
   \author{C. Pin\c con\inst{1,2,3}, M. J. Goupil\inst{3} \and K. Belkacem\inst{3}
%\fnmsep\thanks{Just to show the usage of the elements in the author field}
          }
          
  \authorrunning{C. Pin\c con et al.}
   \institute{
   	STAR Institute, Université de Liège, 19C Allée du 6 Août, B-4000 Liège, Belgium\\
   	 \email{charly.pincon@uliege.be}
   	 \and
   	 Institut d’Astrophysique Spatiale, UMR8617, CNRS, Universit\'e Paris XI, B\^atiment 121, 91405 Orsay Cedex, France
         \and
             LESIA, Observatoire de Paris, PSL Research University, CNRS, Universit\'e Pierre et Marie Curie,
   	Universit\'e Paris Diderot,  92195 Meudon, France%\thanks{The university of heaven temporarily does not accept e-mails}
             }

   \date{Received 7 October 2019 / Accepted 6 December 2019 }

  \abstract
   {The space-borne missions CoRoT and {\it Kepler} have already brought stringent constraints on the internal structure of low-mass evolved stars, a large part of which results from the detection of mixed modes. However, all the potential of these oscillation modes as a diagnosis of the stellar interior has not been fully exploited yet. In particular, the coupling factor or the gravity-offset of mixed modes, $q$ and $\varepsilon_{\rm g}$, are expected to provide additional constraints on the mid-layers of red giants, which are located between the hydrogen-burning shell and the neighborhood of the base of the convective zone. The link between these parameters and the properties of this region, nevertheless, still remains to be precisely established.}
   {In the present paper, we investigate the potential of the coupling factor in probing the mid-layer structure of evolved stars.}
   {Guided by typical stellar models and general physical considerations, we modeled the coupling region along with evolution. We subsequently obtained an analytical expression of $q$ based on the asymptotic theory of mixed modes and compared it to observations.
   }
   {We show that the value of $q$ is degenerate with respect to the thickness of the coupling evanescent region and the local density scale height. On the subgiant branch and the beginning of the red giant branch (RGB),  the model predicts that the peak in the observed value of $q$ is necessarily associated with the important shrinking and the subsequent thickening of the coupling region, which is located in the radiative zone at these stages. The large spread in the measurement is interpreted as the result of the high sensitivity of $q$ to the structure properties when the coupling region becomes very thin. Nevertheless, the important degeneracy of $q$ in this regime prevents us from unambiguously concluding on the precise structural origin of the observed values. In later stages, the progressive migration of the coupling region toward the convective zone is expected to result in a slight and smooth decrease in $q$, which is in agreement with observations. At one point just before the end of the first-dredge up and the luminosity bump, the coupling region becomes entirely located in the convective region and its continuous thickening is shown to be responsible for the observed decrease in $q$. We demonstrate that $q$ has the promising potential to probe the migration of the base of the convective region as well as convective extra-mixing during this stage. We also show that the frequency-dependence of $q$ cannot be neglected in the oscillation spectra of such evolved RGB stars, which is in contrast with what is assumed in the current measurement methods. This fact can have an influence on the physical interpretation of the observed values. In red clump stars, in which the coupling regions are very thin and located in the radiative zone, the small variations and spread observed in $q$ suggest that their mid-layer structure is very stable.
   }
   {A structural interpretation of the global observed variations in $q$ was obtained and the potential of this parameter in probing the dynamics of the mid-layer properties of red giants is highlighted. This analytical study paves the way for a more quantitative exploration of the link of $q$ with the internal properties of evolved stars using stellar models for a proper interpretation of the observations. This will be undertaken in the following papers of this series.
   }

   \keywords{asteroseismology -- stars: oscillations -- stars: interiors -- stars: evolution
               }

   \maketitle

%%%%%%%%%%%%%%%%%%%%

%_________________________________
%
\section{Introduction}
%_________________________________
\label{introduction}

Over the last decades, asteroseismology has been proven to be a powerful method to probe the interior of a large variety of stars at diverse evolutionary stages \citep[e.g.,][and references therein]{DiMauro2016}. Among different classes of pulsations are the solar-like oscillations, which are global stable modes that are both excited and damped by turbulence in the outermost convective layers of stars. The frequency pattern of these modes is known to be an important source of constraints on stellar properties \citep[e.g.,][]{Samadi2015}. However, because of intrinsically low amplitudes, the observation of such oscillations in distant stars were difficult before the advent of the photometric space-borne missions CoRoT \citep{Baglin2006a,Baglin2006b} and {\it Kepler} \citep{Borucki2010}. Indeed, both missions marked a key turning point in the field. They provided high-quality seismic data that enabled the detection of solar-like oscillations in thousands of low-mass stars from the main sequence to the red giant phases. This large number of constraints subsequently led to a revolution in our understanding of stellar structure and stellar evolution \citep[see][for a review]{Chaplin2013}.

Among the main successes of the aforementioned missions was the detection of mixed modes in red giant and red clump stars \citep[e.g.,][]{Bedding2010,Beck2011,Mosser2011}. Mixed modes can propagate in two resonant cavities located in the inner radiative zone, where they behave as gravity modes, and in the external envelope, where they behave as pressure modes. Both cavities are coupled by an intermediate region located between the profiles of the Brunt-Väisälä and Lamb frequencies where mixed modes are evanescent \citep[e.g.,][for a recent review; see also \figurename{}~\ref{propdiag}]{Hekker2017}. After the exhaustion of the central hydrogen at the end of the main sequence, the structural changes resulting from the contraction of the helium core and the expansion of the envelope in evolved stars made such a coupling possible. The frequency pattern of mixed modes observed in these stars contains the signature of the innermost layers in which they propagate and thus has the potential to provide stringent constraints on their properties.

First, the detection of mixed modes gave us access to the period spacing of dipolar modes, denoted with $\Delta \Pi_1$, which was measured in about 6100 evolved stars observed by the {\it Kepler} satellite \citep[e.g.,][]{Mosser2012a,Vrard2016}. This seismic index was shown to be correlated with the mass of the helium core \citep[e.g.,][]{Montalban2013}. Besides, the frequency large separation between two consecutive radial modes, denoted with $\Delta \nu$, is sensitive to the mean envelope density; together, they form a couple of indicators of the evolutionary state during the post-main sequence that permits to unambiguously discriminate between hydrogen-burning shell stars on the red giant branch (hereafter, RGB) and helium-burning core stars on the red clump \citep[e.g.,][]{Bedding2011,Mosser2012a,Mosser2014}. Second, the measurement of rotationally-induced frequency splittings provided an estimate of the core rotation rate in hundreds of evolved stars \citep[e.g.,][and references therein]{Gehan2018}
that turned out to be much lower than predicted in current stellar models \citep[e.g.,][]{Marques2013,Cantiello2014}.
These seismic observations represent a goldmine of information to identify and characterize the missing mechanism that is responsible for these slow core rotations among several candidates \citep[e.g.,][and references therein]{Belkacem2015b,Belkacem2015a,Spada2016,Pincon2017}. 

All the seismic constraints provided by mixed modes resulted not only in a better characterization of the interior of evolved stars themselves, but also of the structural properties of their past and future evolutionary stages \citep[e.g.,][for such investigations]{Montalban2013,Cantiello2014}. Moreover, such advancements on a large sample of stars benefited to other connected fields as Galactic archaeology, for which evolved stars represent key targets owing to their high luminosity and their wide mass range \citep[e.g.,][]{Miglio2013,Mosser2016,Noels2016}.
However, all the probing potential of mixed modes has not been explored yet. In particular, the gravity offset and coupling factor of mixed modes, denoted with $\varepsilon_{\rm g}$ and $q$, respectively, are expected to bring additional constraints on the red giant structure. These seismic parameters, which characterize the frequency pattern of mixed modes, were measured in a large sample of stars by \cite{Mosser2017b,Mosser2018}. The variations in the gravity offset during the evolution on the RGB were recently shown by \cite{Pincon2019} to be sensitive to the density contrast between the core and the envelope, and to the location of the base of the convective region. Regarding the coupling factor, the relation between the observed values and the internal properties nonetheless still calls for further investigations.

The coupling factor describes the degree of interaction between the central and the external resonant cavities of mixed modes. The theoretical link between $q$ and the internal structure was originally provided by the asymptotic analysis of mixed modes of \cite{Shibahashi1979} in the limiting case of a thick intermediate evanescent zone (sometimes called the weak coupling hypothesis). This study showed that $q$ depends on the wave transmission coefficient through the evanescent zone, which is located between the hydrogen-burning shell and the neighborhood of the base of the convective zone. This was confirmed later by \cite{Takata2016a,Takata2016b} from a more general point of view. While $\Delta \nu$ and $\Delta \Pi_1$ are sensitive to the external layers and the surrounding of the helium core, respectively, the coupling factor thus promises to bring us substantial information on this intermediate region, where most of the star luminosity is produced at this evolutionary stage and where mixing processes at the interface between the convective and radiative zones can occur.

The first measurements of $q$ in {\it Kepler} evolved stars were obtained by fitting the asymptotic expression derived by \cite{Shibahashi1979} to the observed frequency pattern of dipolar mixed modes, considering $q$ as a free parameter \citep[e.g.,][]{Mosser2012a,Buysschaert2016,Vrard2016}. All these works demonstrated that this asymptotic analytical form well reproduces the observed frequency pattern but requires values of $q$ higher than $1/4$ in some stars, in disagreement with the analytical predictions obtained in the weak coupling hypothesis. \cite{Jiang2014} reached the same conclusions by comparing the asymptotic frequencies with the theoretical frequencies computed via an oscillation code in evolved stellar models \citep[see also][]{Hekker2018}. These works underlined the need for going beyond the weak coupling hypothesis, which is a prerequisite for fully grasping the information brought by $q$ on the internal structure. Motivated by such discrepancies, \cite{Takata2016a} then developed an asymptotic analysis of mixed modes in the reverse limiting case of a very thin evanescent zone (sometimes called the strong coupling hypothesis). In this formalism, the expression of mixed modes is similar to that in the weak coupling hypothesis, except that $q$ can take values between zero and unity. This result validated the fitting procedures used in the previous observational studies, which assumed $q$ as a free parameter between zero and unity, as well as their outcomes.

In parallel to the work of \cite{Takata2016a}, \cite{Mosser2017b} proposed a new automated method to measure $q$ in a large set of stars. This in-depth investigation confirmed previous observational results and provided a coherent view of the coupling factor for dipolar modes during the post-main sequence evolution. The large-scale measurement in more than 5000 evolved stars showed that the value of $q$ changes during evolution with clear signatures at the transition from the subgiant phase to the RGB as well as to the red clump, where high values of $q$ are observed, in qualitative agreement with the recently-proposed strong coupling formalism. Assuming the evanescent region is located in the radiative layers just above the helium core, as suggested in stellar models of young red giant stars, \cite{Mosser2017b} showed that the value of $q$ depends both on the variation scale height of the Brunt-Väisälä  frequency and on the thickness of the evanescent region; however, they could not directly disentangle the origin of the observed variations in $q$ among these two properties.

In this series of papers, we aim to clarify further the link between the coupling factor of mixed modes and the internal properties of evolved stars from a theoretical point of view. As a stepping stone to the investigation, an analytical approach is considered in this first paper. Using the available asymptotic analyses (i.e., both in the weak and the strong coupling hypotheses), we search for simplified expressions of $q$ and scrutinize in detail the physical information encapsulated inside this parameter and its variations during evolution. The paper is organized as follows. In Sects.~\ref{dipolar} and \ref{basic coupling}, the theoretical background about mixed modes is introduced within the framework of the asymptotic limit, with emphasis on the coupling factor and its physical meaning. Following this introductory material, a simple model is then developed in \sectionname{}~\ref{simple model}. It is based on the power-law behavior of the Brunt-Väisälä and Lamb frequencies in the evanescent region, as assumed by \cite{Mosser2017b}, but in contrast also accounts for the migration of the evanescent region from the radiative to the convective zones during the evolution on the RGB. The relation between $q$ and the properties of the intermediate evanescent region is subsequently examined. The observations are then interpreted in the light of the analytical model in \sectionname{}~\ref{interpretation}. The results and main assumptions used in this work are discussed in \sectionname{}~\ref{discussion} and the conclusions are formulated in \sectionname{}~\ref{conclusions}.

%_________________________________
%
\section{Mixed mode cavities in evolved stars}
%_________________________________

\label{dipolar}

We introduce the basic background about mixed modes in evolved stars within the asymptotic framework, which is in the short-wavelength WKB approximation \citep[e.g.,][]{Gough2007}. The structures of the resonant cavities and the coupling evanescent region are presented. In the following, we focus on dipolar $\ell=1$ modes, with $\ell $ the angular degree, because of the peculiar role they play from an observational point of view. Indeed, among mixed modes whose energy is mainly trapped in the core of evolved stars and that are thus mostly sensitive to the properties of the innermost layers, the dipolar ones are the most easily detectable \citep[e.g.,][]{Dupret2009,Benomar2014,Grosjean2014}. As a result, the measurement of the coupling factor in a large set of stars is currently mainly limited to dipolar modes with {\it Kepler} data. 

%------------------------------
\subsection{Dispersion relation}
%-----------------------------
\label{dispersion}

Solar-like oscillation modes result from the constructive interferences of gravito-acoustic waves stochastically excited in the uppermost layers of stars and traveling back-and-forth several times between the center and the surface. The propagation of these waves is physically ruled by a fourth-order linear differential equation under the adiabatic approximation \citep[e.g.,][]{Unno1989}. In order to analytically study stellar oscillations, the Cowling approximation \citep[e.g.,][]{Cowling1941}, which consists in neglecting the perturbation of the gravitational potential denoted with $\Phi^\prime$ in the following, was usually adopted in previous works. This reduces the adiabatic oscillation equations from the fourth to the second order, and thus makes these tractable with usual asymptotic methods. However, this hypothesis remains questionable for dipolar modes because of their very large horizontal wavelength (see \sectionname{}~\ref{Cowling A}).
Fortunately, owing to the Hamiltonian structure of the adiabatic stellar oscillations and the existence of a general first integral for dipolar modes, \cite{Takata2005,Takata2006b,Takata2006a} demonstrated that the dipolar wave equation can be rewritten in the form of a second-order differential equation, which has the advantage of fully considering the perturbation of the gravitational potential and being analytically tractable. To do this, the dependent variables of the governing equations have to be modified from the standard mass displacement to the reduced displacement, which means the displacement of mass elements minus that of the center of mass of the corresponding concentric mass.

In the asymptotic limit, \cite{Takata2006b,Takata2016a} demonstrated that accounting for the perturbation of the gravitational potential modifies the dispersion relation of dipolar modes; in this case, this takes the form of
\algn{
k_r^2=\frac{\sigma^2}{c^2}\left(\frac{\mathcal{N}^2}{\sigma^2}-1 \right) \left(\frac{\mathcal{S}_1^2}{\sigma^2}-1 \right) \; ,
\label{k_r}
}
where $k_r$ is the local radial wavenumber, $\sigma$ is the angular mode frequency, $c^2=(\Gamma_1 p) /\rho$ is the squared sound speed, $\Gamma_1$ is the first adiabatic index, and $p$ and $\rho$ are the equilibrium pressure and density, respectively. In this latter equation, $\mathcal{N}$ and $\mathcal{S}_1$ are the modified Brunt-Väisälä and Lamb frequencies (for $\ell=1$), respectively, which account for the perturbation of the gravitational potential, and are defined as
\algn{
\mathcal{N}=\frac{N}{J}~~~~\mbox{and }~~~~\mathcal{S}_1=S_1 J \; .
\label{NSJ}
}
where $N$ and $S_\ell$ are the usual Brunt-Väisälä and Lamb frequencies defined as \citep[e.g.,][]{Unno1989}
\algn{
S_\ell^2&=\frac{\ell(\ell+1) c^2}{r^2}\label{Lamb}\\
N^2&=\frac{g}{r}\left(\frac{1}{\Gamma_1}\deriv{\ln p}{\ln r} - \deriv{\ln \rho}{\ln r}\right) \label{BVaisala}\; ,
}
in which $r$ is the radius in the star and $g$ is the gravitational acceleration. 
The $J$ factor in \eq{NSJ} is equal to
\algn{
J=1-\frac{1}{3} \deriv{\ln M_r}{\ln r} = 1-\frac{4\pi r^3 \rho}{3M_r}=1-\frac{\rho(r)}{\rho_{\rm av}(r)} \; ,
\label{def J}
}
in which $M_r$ and $\rho_{\rm av}(r)$ represent the mass and mean density inside the spherical shell of radius $r$, respectively. The continuity equation $(\dd M_r / \dd r) = 4\pi r^2 \rho$ has been used in the second equality of \eq{def J}. 
Under the Cowling approximation, the dispersion relation has the same form as \eq{k_r}, except that $\mathcal{N}$ and $\mathcal{S}_1$ are replaced by the usual critical frequencies $N$ and $S_1$, respectively \citep[e.g.,][]{Unno1989}. Indeed, the adiabatic oscillation equations for dipolar modes obtained within the Cowling approximation can be retrieved from the set of Eqs.~(1)-(9) provided by \cite{Takata2016a} in the non-Cowling case, in the limits of $J\rightarrow 1$ and $\Phi^\prime \rightarrow 0$.

As an illustration, the modified Brunt-Väisälä and $\ell =1$ Lamb frequencies are displayed in \figurename{}~\ref{propdiag} as a function of radius for three 1.2$M_\odot$ stellar models. These models correspond to stars on the subgiant branch (SG), at the beginning of the ascent of the RGB (YR) and just before the bump of luminosity (ER). The modified Brunt-Väisälä and Lamb frequencies are compared to the usual expressions given in \eq{Lamb}{BVaisala}~in the same figure. The internal radiative zone and the adiabatic external convective zone, where $N^2 \approx 0$, can be easily distinguished. To be complete, the $J$ factor is also plotted for the three considered models in \figurename{}~\ref{J graph}. The stellar models were computed with the CESTAM evolution code \citep[][]{Marques2013}, considering a solar mixture following \cite{AGS2009}, solar calibrated initial helium and metal abundances $Y_0=0.25$ and $Z_0=0.013$, and a solar calibrated mixing-length parameter $\alpha_{\rm MLT}=1.65$. The input physics is standard, without microscopic diffusion, rotation nor overshooting. The models and their evolutionary tracks are indicated in the Hertzsprung-Russell diagram (HR diagram) in \figurename{}~\ref{HR}.
As shown in \figurename{}~\ref{propdiag}, a clear difference can be observed between the usual and the modified critical frequencies below the hydrogen-burning shell, because of very low values of $J$. In the helium core, the density is nearly uniform and we can show by a second-order Taylor expansion near the center that $J\approx (-7 \rho_{\rm c}^{\prime \prime}/10 \rho_{\rm c}) ~r^2$, with $\rho_{\rm c}$ and $\rho_{\rm c}^{\prime\prime}$ the density and its second-derivative with respect to radius at the center. The effect of the Cowling approximation on the behavior of dipolar mixed modes is thus important in the central layers. In contrast, the usual and modified critical frequencies share comparable behavior above the hydrogen-burning shell (i.e., $N\approx \mathcal{N}$ and $S_1\approx \mathcal{S}_1$). However, \cite{Jiang2014} and \cite{Mosser2017b} showed using stellar models that the Cowling approximation can lead to incorrect estimates of the mode coupling through the evanescent region. We therefore keep the influence of the perturbation of the gravitational potential in the following by considering the modified expression of the critical frequencies. The effect of the Cowling approximation on the mode coupling is discussed in more detail in \sectionname{}~\ref{weak coupling}, after having properly introduced the coupling factor in the asymptotic framework.

%-------------------------------
\subsection{Propagation diagram in evolved stars}
%-------------------------------
\label{Prop diag}
\begin{figure}
\centering
\includegraphics[width=\hsize,trim= 0.8cm 0cm 0.8cm 1cm, clip]{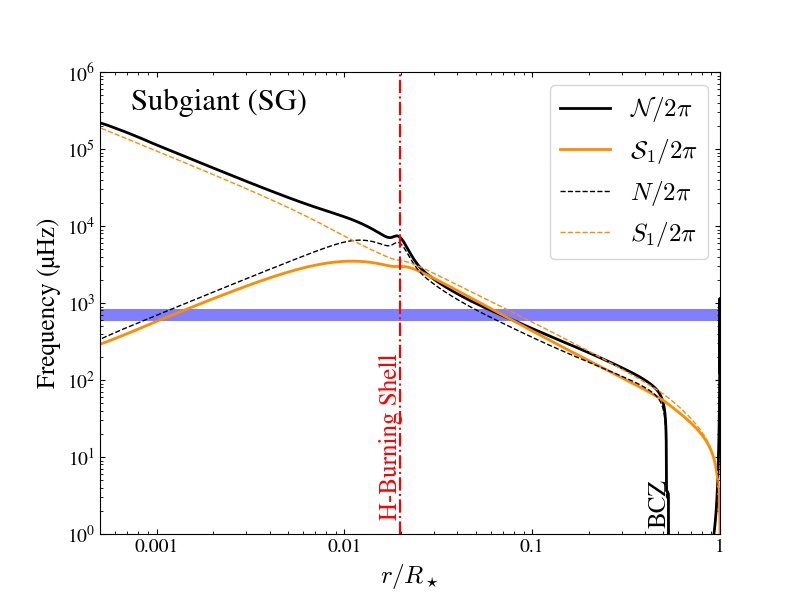} 
\includegraphics[width=\hsize,trim= 0.8cm 0cm 0.8cm 0cm, clip]{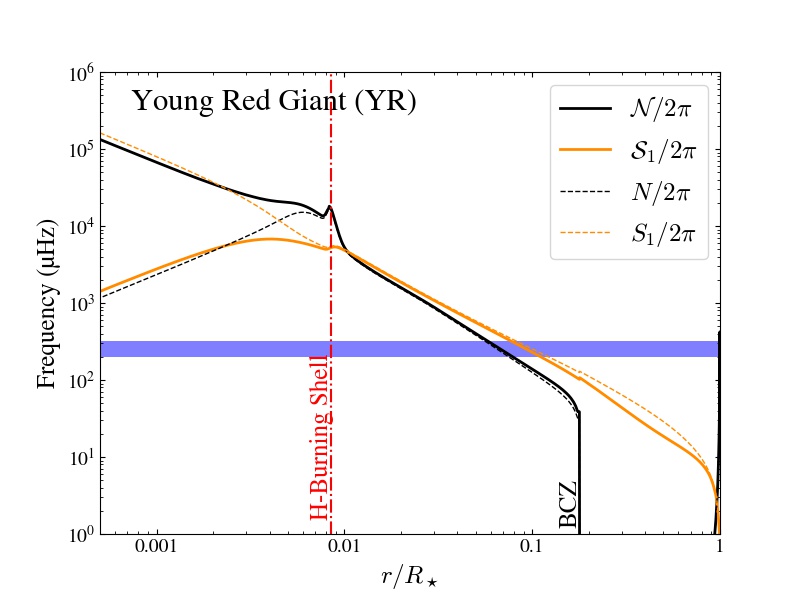} 
\includegraphics[width=\hsize,trim= 0.8cm 0cm 0.8cm 0cm, clip]{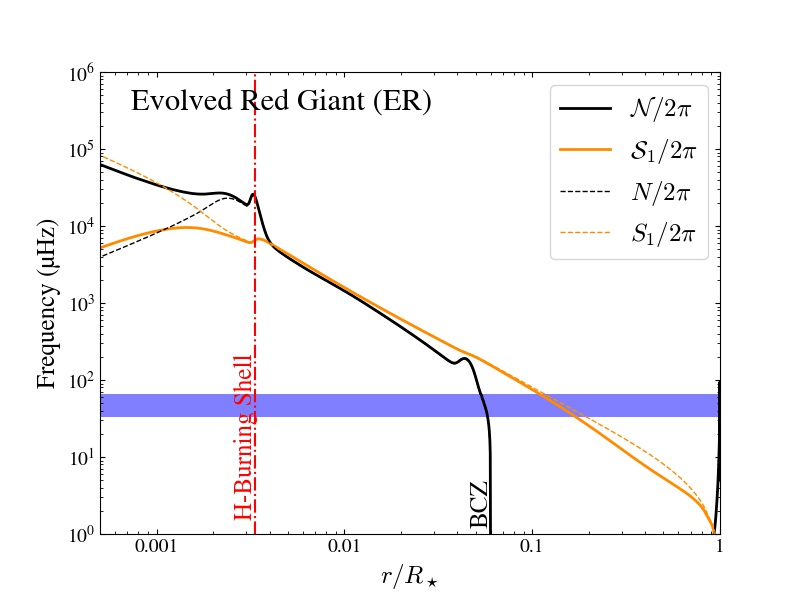} 
\caption{Profiles of the modified (solid lines) and usual (dashed lines) Brunt-Väisälä and Lamb frequencies (for $\ell=1$) as a function of radius in three $1.2M_\odot$ stellar models, corresponding to a subgiant star (SG), a young (YR) and an evolved (ER) red giant stars, respectively (see HR diagram in \figurename{}~\ref{HR}). The base of the convective zone (BCZ) and the hydrogen-burning shell (vertical dash-dotted line) are indicated. The blue horizontal strip symbolizes a 6$\Delta\nu$-wide frequency range around $\nu_{\rm max}$ computed via \protect\eq{nu_max}{Delta nu}. Propagation occurs where \smash{$2\pi \nu\lesssim (\mathcal{N}$ and $\mathcal{S}_1)$} or $2\pi \nu \gtrsim (\mathcal{N}$ and $\mathcal{S}_1)$, and evanescence otherwise.}
\label{propdiag}
\end{figure}

As shown by \eq{k_r}, the propagation and trapping of waves in stars are determined by the profiles of the Brunt-Väisälä and Lamb frequencies, and on the typical oscillation frequency. Excitation models by turbulent convection \citep[e.g.,][]{Samadi2015} and observations \citep[e.g., see Fig.~6 in][]{Mosser2013c} show that solar-like oscillations are efficiently generated in a narrow frequency range around a certain frequency at maximum oscillation power, which is denoted with $\nu_{\rm max}$. Empirical and theoretical studies demonstrated that $\nu_{\rm max}$ scales as the acoustic cut-off frequency at the stellar surface \citep[e.g.,][]{Brown1991,Kjeldsen1995,Belkacem2011,Belkacem2013}. It results in a scaling relation between $\nu_{\rm max}$, the mass $M$, the radius $R$ and the effective temperature $T_{\rm eff}$ of a given star, that reads
\algn{
\nu_{\rm max}\approx \nu_{{\rm max},\odot} \left( \frac{M}{M_\odot}\right)\left( \frac{R}{R_\odot}\right)^{-2}
\left( \frac{T_{\rm eff}}{T_{{\rm eff},\odot}}\right)^{-1/2} \; ,
\label{nu_max}
}
which has been scaled for our purpose to the solar reference value \smash{$\nu_{{\rm max},\odot} = 3104~\mu$Hz} such as prescribed by \cite{Mosser2013b}. Typically, mixed modes are visible over a $6\Delta \nu$-wide frequency range around $\nu_{\rm max}$ \citep[e.g.,][]{Grosjean2014,Mosser2018}. Here, $\Delta \nu$ represents the frequency spacing between two pressure radial modes with consecutive radial orders, the asymptotic expression of which is given by \citep[e.g.,][]{Unno1989}
\algn{
\Delta \nu = \left( 2 \int_0^{R} \frac{\dd r}{c}\right)^{-1} \; .
\label{Delta nu}
}
Such typical observed frequency ranges are represented in \figurename{}~\ref{propdiag}. We find $\nu_{\rm max}\approx 700,260,$ and $50~\mu$Hz and $\Delta \nu\approx 43,20,$ and $6~\mu$Hz for the models SG, YR, and ER, respectively. 

According to \eq{k_r}, propagation diagrams in \figurename{}~\ref{propdiag} show that observed oscillation modes can propagate (i.e., $k_r^2>0$) in a central radiative cavity where \smash{$\sigma^2 < (\mathcal{N}^2 ~\mbox{and}~ \mathcal{S}_1^2)$} and they behave as gravity modes (buoyancy cavity), and in an outer cavity where \smash{$\sigma^2 >  (\mathcal{N}^2~ \mbox{and}~ \mathcal{S}_1^2)$} and they behave as pressure modes (acoustic cavity). In these typical models, we also see that considering an asymptotic description is reasonable since $ \nu_{\rm max}\ll N/2\pi$ and $\nu_{\rm max}\gg S_1/2\pi$ inside both cavities, respectively, so that $k_r \gg 1/r$ according to \eq{k_r} and the number of radial nodes is much larger than unity. Both resonant cavities are coupled by an intermediate evanescent zone (i.e., $k_r^2<0$) located between the profiles of the (modified) Brunt-Väisälä and Lamb frequencies, through which the mode energy is transmitted from a cavity to another cavity. More exactly, the evanescent region is located between the turning points $r_1(\sigma)$ and $r_2(\sigma)$ where $k_r^2$ vanishes, that is, where $\mathcal{N}(r_1)=\sigma$ and $\mathcal{S}_1(r_2)=\sigma$.
These are the so-called mixed modes with a dual character of pressure and gravity modes. Such a configuration of coupling is not encountered in main sequence stars since $\nu_{\rm max}$ is greater than the maximum value of the Brunt-Väisälä frequency in the radiative zone. Nevertheless, as a star evolves on the post-main sequence, its helium core contracts and becomes denser and denser. As observed in \figurename{}~\ref{propdiag}, this leads to an increase in the value of the Brunt-Väisälä frequency near the hydrogen-burning shell, denoted with $N_{\rm M}$, that is, just above the helium core. Indeed, since $N^2\sim g/r$ at first order in the radiative zone, with $g=G M_r /r^2$, we have $N^2_{\rm M} \sim G M_{\rm He} /R_{\rm He}^3 \propto G \bar{\rho}_{\rm He}$ where $M_{\rm He}$, $R_{\rm He}$ and $\bar{\rho}_{\rm He}$ are the mass, radius and mean density of the helium core, respectively. Similarly, using the hydrostatic equilibrium and the sound speed definition, we can show that $S_1^2 \sim g/r$ at first order in this region (see \eq{S_l g/r} for instance); the profile of the Lamb frequency thus also increases during evolution in the inner layers. In contrast to the core contraction, the envelope expands (i.e., $R$ increases) while the stellar luminosity $L$ increases and the effective temperature decreases. From the Stefan-Boltzmann relation, which is $L\propto R^2 T_{\rm eff }^4$, it is thus obvious that the effective temperature $T_{\rm eff}$ decreases more slowly than \smash{$1/\sqrt{R}$}. According to \eq{nu_max} at a given mass, $\nu_{\rm max}$ thus decreases during evolution, as seen in \figurename{}~\ref{propdiag}. As a result, the combined effects of the core contraction and envelope expansion make the coupling between the inner buoyancy cavity and the outer acoustic cavity possible in evolved stars. This happens as soon as the subgiant stage starts (see the transition between the dashed and solid lines in \figurename{}~\ref{HR}).

\begin{figure}
\centering
\includegraphics[width=\hsize,trim= 0.8cm 0cm 0.8cm 1cm, clip]{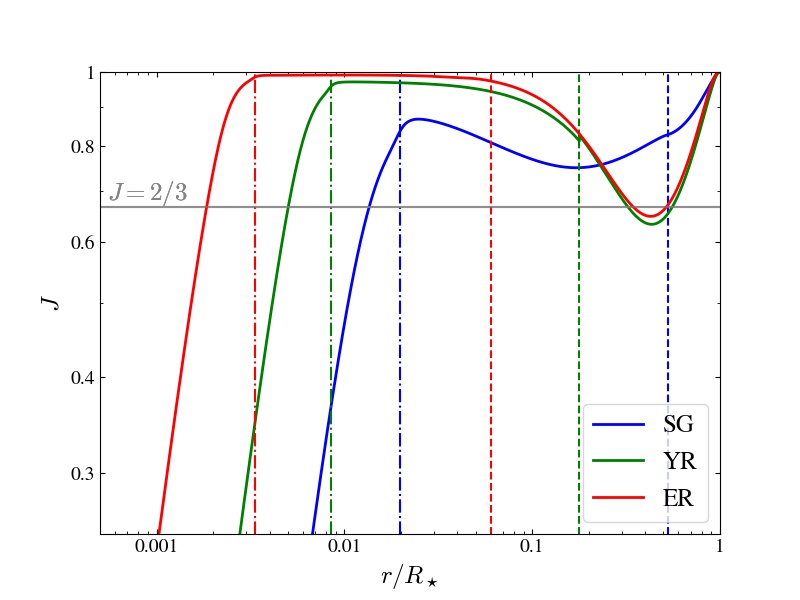} 
\caption{$J$ factor defined in \protect\eq{def J} as a function of radius in the three $1.2M_\odot$ stellar models considered in \figurename{}~\ref{propdiag}. The vertical dash-dotted and dashed lines represent the hydrogen-burning shell and the base of the convective zone, respectively.}
\label{J graph}
\end{figure}

Figure~\ref{propdiag} also shows that the intermediate evanescent zone that couples the buoyancy and acoustic cavities evolves on the post-main sequence. It even radically changes in nature and in structure at some point during the ascent on the RGB.
Indeed, for subgiant stars (e.g., model SG) and young red giant stars (e.g., model YR), the evanescent zone is located inside the radiative zone, between the hydrogen-burning-shell and the base of the convective zone. In the following, we call such a coupling zone {\bf Type-a evanescent region}. As shown in \figurename{}~\ref{propdiag}, the radial extent of Type-a evanescent regions can be very thin on the subgiant branch (e.g., model SG), and becomes slightly thicker in young red giant stars (e.g., model YR). At one point during the evolution, the value of $\nu_{\rm max}$ becomes low enough for the evanescent region to be entirely located in the convective zone and its lower bound to be very close to the radius at the base of the convective zone, denoted with $r_{\rm b}$ (e.g., model ER). In the following, we call such a coupling zone {\bf Type-b evanescent region}. As shown in \figurename{}~\ref{propdiag}, Type-b evanescent regions are expected to be much thicker than Type-a evanescent regions. The transition phase between both configurations is indicated by a red square in the HR diagram of \figurename{}~\ref{HR} and turns out to occur before the luminosity bump. 
In red clump stars, the structural readjustments resulting from the helium burning ignition lead again to Type-a evanescent regions \citep[e.g., see  Fig.~9 in][]{Hekker2018}.

We note that Type-a and Type-b evanescent regions generally match the cases $a$ and $b$ considered in \cite{Pincon2019}, which are associated with stars such as $\nu_{\rm max} \gtrsim \mathcal{N}_{\rm b}/2\pi $ and $\nu_{\rm max} \lesssim \mathcal{N}_{\rm b}/2\pi $, respectively, where $\mathcal{N}_{\rm b}$ is the value of the modified Brunt-Väisälä frequency just below the base of the convective region. Using standard stellar models with the same input physics as used to build the stellar models considered in \figurename{}~\ref{propdiag}, \cite{Pincon2019} estimated that the transition from case $a$ to case $b$, thus corresponding to the transition from Type-a to Type-b evanescent regions, occurs for values of  $\nu_{\rm max}$ equal to \smash{$\nu_{\rm max, t}\approx 110$, 95, 80, 70, 60, and 50$~\mu$Hz,} for typical stellar masses of $1$, $1.2$, $1.4$, $1.6$, $1.8$, and $2M_\odot$, respectively. Using the $\nu_{\rm max}$-$\Delta\nu$ scaling law $ \Delta \nu\approx 0.28 \nu_{\rm max}^{0.75}$ with $\Delta \nu$ and $\nu_{\rm max}$ expressed in $\mu$Hz \citep[e.g.,][]{Mosser2012b,Peralta2018}, this transition is expected to occur for values of $\Delta \nu$ between $10~\mu$Hz and $5~\mu$Hz in the same mass range. This change of configuration during evolution impacts the coupling of mixed modes through the intermediate evanescent region and therefore needs to be considered.

\begin{figure}
\centering
\includegraphics[width=\hsize,trim= 0.8cm 0cm 0.8cm 1cm, clip]{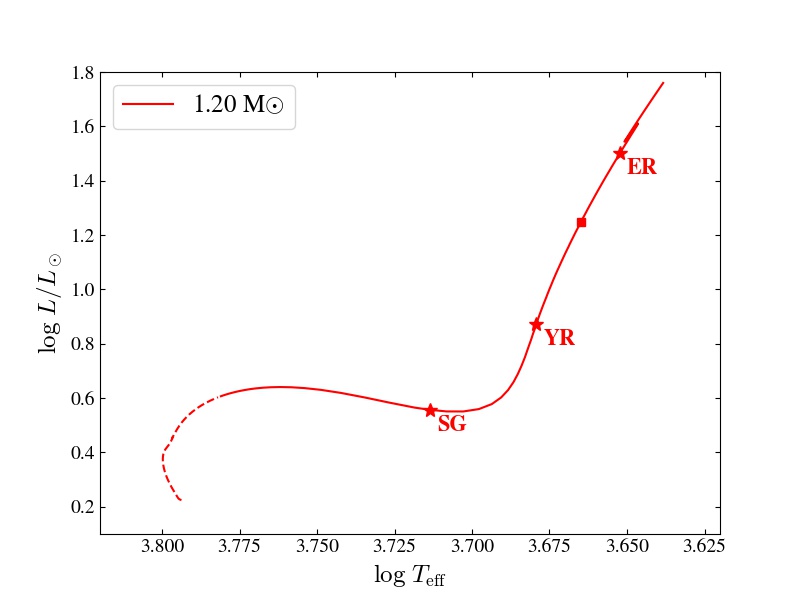} 
\caption{Evolutionary sequence of 1.2$M_\odot$ stellar models in the HR diagram, from the zero-age main sequence to the tip of the RGB. The solid line represents the domain where a coupling between the core and the envelope is possible and mixed modes can be observed, unlike the dashed line. The red stars represent the three models (SG, YR and ER) considered as examples in \figurename{}~\ref{propdiag}. The square qualitatively labels the transition point from a Type-a evanescent region to a Type-b evanescent region.}
\label{HR}
\end{figure}
%

%----------------------------------
\subsection{Structure of the evanescent coupling region}
%-----------------------------------
\label{struct ev}

The actual shape of the evanescent zone, and thus the coupling between the resonant cavities, depend on the profiles of the (modified) Brunt-Väisälä and Lamb frequencies since, according to \eq{k_r}, these latter define the decay length of the mode amplitude in this region. Actually, as shown in \appendixname{}~\ref{equivalence relation}, both $\mathcal{N}$ and $\mathcal{S}_1$ profiles are intrinsically related to the mass distribution inside the star, represented by the $J$ factor in \eq{def J}.

In the radiative zone, between the hydrogen-burning shell and the base of the convective region, \figurename{}~\ref{J graph} shows that $J$ remains nearly uniform with values higher than $0.7$ in the considered models, indicating a high density contrast between the considered layers and the helium core. Using the internal structure equations and simple analytical models, we demonstrate in \appendixname{}~\ref{upper rad} that this high density contrast is the main responsible for the similar power-law behavior of $\mathcal{N}$ and $\mathcal{S}_1$ with respect to radius in this region (see \figurename{}~\ref{propdiag}). Actually, the parallel profiles of $\mathcal{N}$ and $\mathcal{S}_1$ result from their dependence in $(g/r)^{1/2}$, as shown through \eq{def A and V}{SVNA}~considering $(\dd \ln p/\dd \ln r)$, $(\dd \ln \rho/\dd \ln r)$ and $J$ as nearly uniform. Since by definition $\mathcal{N}(r_2)=\sigma$ and $\mathcal{S}_1(r_1)=\sigma$, we thus can write in a good approximation in the upper radiative layers \citep[e.g.,][]{Takata2016a,Mosser2017b}
\algn{
\mathcal{N} (r) \approx \sigma \left( \frac{r_1}{r}\right)^{\beta}~~~~\mbox{and}~~~~\mathcal{S}_1(r) \approx \sigma\left( \frac{r_2}{r}\right)^{\beta} \; ,
\label{power law N and S}
}
where $\beta$ is defined as the (minus) logarithmic derivative of $\mathcal{S}_1$ in the middle of the evanescent region at $r_{\rm ev}$, that is,
\algn{
\beta\equiv-\left(\deriv{\ln \mathcal{S}_1}{\ln r} \right)_{r_{\rm ev}}\; .
}
%. 
In this power-law configuration, a simple relation exists between $\beta$, $J$ and the local polytropic exponent, $\gamma=(\dd \ln p /\dd \ln \rho)$, which reads
(see \appendixname{}~\ref{equivalence relation})
\algn{
\beta \approx\frac{3}{2}J\approx\frac{1}{2-\gamma} \; .
\label{poly beta}
}
Considering both limiting cases of an isothermal perfect gas or a relativistic photon gas, we can expect the values of $\gamma$ to range between about 1 and $4/3$ in the radiative layers surrounding the helium core. According to \eq{poly beta}, we thus expect \smash{$J$} to range between about $2/3$ and 1, and $\beta$ to range between about 1 and 1.5. We check that such an expected range for $\beta$ is in agreement with stellar models of typical subgiant and RGB stars. During this phase of the evolution, the density contrast between the core and the envelope increases, and thus the value of $J$ increases too (see \figurename{}~\ref{J graph}). According to \eq{poly beta}, the value of $\beta$ is therefore also expected to increase. This trend in the slope of $\mathcal{N}$ and $\mathcal{S}_1$ is clearly observed in stellar models, as shown in \figurename{}~\ref{propdiag}. 

In the deep layers of the convective region of evolved red giant stars (e.g., model ER), the equation of state is adiabatic and \figurename{}~\ref{propdiag} shows that the Lamb frequency closely behaves as a power law of radius, so that we can write at first approximation
\algn{
\mathcal{N}\approx 0~~~~\mbox{and}~~~~\mathcal{S}_1(r) \approx \sigma\left( \frac{r_2}{r}\right)^{\beta} \; .
\label{model B}
}
The inner turning point associated with the evanescent region is assumed to coincide with the radius at the base of the convective zone, $r_{\rm b}$. Although \eq{model B} represents a first-step description of the evanescent region, we note that the hydrostatic equilibrium and continuity equations do not allow $\mathcal{S}_1$ to rigorously follow a power law of radius in such an adiabatic region (see \appendixname{}~\ref{equivalence in conv} for details). In particular, this implies that the same relation as in \eq{poly beta} does not apply in this region; instead, the link between $\beta$, $J$ and $\gamma$ is ruled by a more complicated differential equation. Using typical stellar models of evolved red giant stars with Type-b evanescent regions, we find that the value of $\beta$ lies in a confident range going from 1.2 to 1.5 over the evanescent region, and that the value of $J$ ranges between 0.8 and 1.

In the following, the approximations made in \eq{power law N and S}{poly beta}~or in \eq{model B} enable us to analytically express $q$ in Type-a or Type-b evanescent regions, respectively, while accounting for the relation between these three variables in a consistent way. Before going further, we first need to introduce in more detail the coupling factor of mixed modes.

%_________________________________________________
%
\section{Theoretical background about the coupling factor}
%_________________________________________________
\label{basic coupling}

In this section, the physical meaning of the coupling factor of mixed modes is introduced from a general point of view and discussed by means of an instructive toy model. Its general expressions in the asymptotic limit are subsequently presented.

%--------------------------------
\subsection{Physical definition}
%--------------------------------
\label{resonance condition}

The eigenfrequencies of mixed modes are selected by the boundary conditions that are imposed at the center and surface, resulting in a discrete spectrum. The asymptotic resonance condition was first provided by the analyses of \cite{Shibahashi1979} and \cite{Tassoul1980} in the hypothesis of a very thick evanescent coupling region. This expression was rewritten in an observational context by \cite{Mosser2012a} in the form of
\algn{
\cot\Theta_{\rm g} \tan \Theta_{\rm p}= q \; ,
\label{resonance}
}
where $\Theta_{\rm g}$ and $\Theta_{\rm p}$ are frequency-dependent phase terms associated with the propagation of waves in the inner and external cavities, respectively, and $q$ is the so-called coupling factor. The observed frequency pattern of mixed modes in stars is thus partly determined by the value of the coupling factor \citep[e.g., see][]{Mosser2012a,Mosser2017b}. 
Using basic wave principles and the WKB approximation, \cite{Takata2016a,Takata2016b} later demonstrated that the form of the resonance condition in \eq{resonance} actually holds true whatever the thickness of the evanescent region. The author showed in general terms that $\Theta_{\rm g}$ and $\Theta_{\rm p}$ are both equal to the sum of the wavenumber integral over each cavity and of the phase lags introduced at the wave reflection near their boundaries (i.e., the associated turning points), and that $q$ is expressed by
\algn{q=\frac{1-\sqrt{1- T^2}}{1+\sqrt{1-T^2}} \; ,
\label{def q}
}
where $T^2$ is the transmission coefficient of the wave energy through the intermediate evanescent zone. The value of $q$ must therefore depend on the properties of this region.

From \eq{resonance}{def q}, we understand that the coupling factor measures the degree of interaction between both cavities, with values between $0$ and $1$. For example, in the limit of $q\rightarrow 0$ ($T^2\rightarrow 0$), the cavities are independent: from \eq{resonance}, either $\cos \Theta_{\rm g}=0$ or $\sin \Theta_{\rm p}=0$. These conditions are equivalent to the Borh-Sommerfeld's quantization condition for a gravity mode trapped in the inner cavity\footnote{An essential difference exists with the resonance condition of a pure gravity mode found by \cite{Shibahashi1979}. Indeed, in our notation, Eq.~(22) of his paper instead gives $\sin \Theta_{\rm g} = 0$. Actually, the difference comes from the fact that \cite{Shibahashi1979} considered that the outer turning point $r_1$ is defined as the radius where $\sigma=S_1(r_1)$ instead of where $\sigma=N(r_1)$ in our case. This change in the nature of the turning point therefore results in a phase shift of $\pi/2$ in $\Theta_{\rm g}$ between both cases, in a similar way to the effect of the Cowling approximation \cite[see Sect.~4.2 in][]{Takata2016a}.} or for a pressure mode trapped in the envelope, respectively. In contrast, in the limit of $q\rightarrow 1$ ($T^2\rightarrow~1$), there is no evanescent zone but a unique resonant cavity. In the intermediate case, \cite{Takata2016b} showed that the value of $q$ rules the squared ratio of the mode amplitude in the acoustic cavity to the mode amplitude in the buoyancy cavity. Actually, \citet[][cf. constants $a$ and $c$ in \appendixname{}~A.3 of this paper]{Goupil2013} demonstrated that this ratio is equivalent to the ratio of the mean time-averaged mode kinetic energy over one wavelength in the acoustic cavity, denoted with $\mathcal{E}_{\rm p}$, to that in the buoyancy cavity, denoted with $\mathcal{E}_{\rm g}$. From Eq~(22) of \cite{Takata2016b}, it can be rewritten
\begin{equation}
\frac{\mathcal{E}_{\rm p}}{\mathcal{E}_{\rm g}} = \frac{\sin ( 2 \Theta_{\rm g} )}{\sin(2 \Theta_{\rm p})} =\frac{1}{q} \frac{\cos ^2( \Theta_{\rm g} )}{\cos^2( \Theta_{\rm p})} = q  \frac{\sin^2 (  \Theta_{\rm g} )}{\sin^2( \Theta_{\rm p})} \; ,
\label{E ratio}
\end{equation}
where both last equations result from the use of \eq{resonance}.
For gravity-dominated mixed mode, which are mixed modes with a gravity-like behavior such as $\cos \Theta_{\rm g}\approx0$ and $\cos\Theta_{\rm p}\approx 0$, $\mathcal{E}_{\rm g}$ in the inner cavity is higher than $\mathcal{E}_{\rm p}$ in the envelope by a factor of $1/q$. Inversely, for pressure-dominated mixed modes, which are mixed modes with a pressure-like behavior such as $\sin \Theta_{\rm p}\approx0$ and $\sin \Theta_{\rm g}\approx 0$, $\mathcal{E}_{\rm g}$ is smaller than $\mathcal{E}_{\rm p}$ by a factor of $q$. In other cases depending on the mode frequency, the ratio vary between these two extrema, that is, $q \le (\mathcal{E}_{\rm p}/\mathcal{E}_{\rm g}) \le 1/q$. From this latter inequality, it is obvious that the coupling factor determines how the mode energy is distributed between the inner cavity and the envelope, and thus the character of the mode. Indeed, small values of $q$ preferentially result in eigenmodes with a dominant character (i.e., pressure-dominated or gravity-dominated modes), while values of $q$ close to unity result in eigenmodes with an energy more uniformly distributed between both cavities and thus a pronounced pressure-gravity duality. The value of the coupling factor therefore plays an important role in the detectability of the mixed mode frequency pattern \citep[e.g.,][]{ Grosjean2014,Mosser2018}. In order to examine the potential of the coupling factor as a seismic diagnosis, we need to understand how the value of $q$ depends on the properties of the evanescent region.

%--------------------------------
\subsection{An instructive toy model: a parabolic evanescent zone}
%-------------------------------
\label{toy model}

Equation~(\ref{def q}) shows that $q$ depends only on the wave transmission factor through the intermediate evanescent zone. While this formulation is very simple, it is much less easy in practice to grasp how an incident wave behaves as it encounters such a barrier during its travel and how the wave transmission depends on the structure of the barrier. As a first step, it is therefore instructive to study the wave transmission-reflection mechanism in simplified cases that are analytically tractable and can give us qualitative clues on this physical process. 
The most simple situation is to consider a parabolic evanescent region: one assumes that the squared local radial wavenumber is represented by a second-degree polynomial. This is actually similar to study the propagation of a particle in an inverted harmonic potential in quantum physics \citep[e.g.,][]{Heim2013}.

In such a toy model, the wave equation for a given wave function $\Psi$ reads
\algn{
\derivs{\Psi}{r}+k_r^2\Psi=0 ~~~~{\rm with}~~~~ k_r^2= \left\vert k_r(r_{\rm ev})^2 \right\vert\left[\frac{4\left(r-r_{\rm ev}\right)^2}{\Delta r^2}-1\right] \; ,
\label{Psi eq}
}
where $r_{\rm ev}$ is the radius at the middle of the evanescent zone and $\Delta r$ is its radial extent.
The analytical derivation of the wave energy transmission coefficient through such an evanescent zone is detailed in \appendixname{}~\ref{parabolic barrier model} following the work of \cite{Phinney1970}. Its final expression reads
\algn{
T^2=\frac{1}{1+e^{2 \mathcal{I}}}\; , \label{T parabolic}
}
where $\mathcal{I}$ corresponds to the wavenumber integral over the evanescent region, the general definition of which is provided by
\algn{
\mathcal{I}\equiv\int_{r_{\rm ev}-\Delta r/2}^{r_{\rm ev}+\Delta r/2} \vert k_r\vert~\dd r  \; .
\label{k_r integral}
}
The lower and upper bounds of the integral in \eq{k_r integral} correspond to the  turning points $r_1$ and $r_2$ defined in \sectionname{}~\ref{Prop diag}, respectively.
In the case of a parabolic evanescent zone, we find according to \eq{Psi eq} that
\algn{
\mathcal{I}= \frac{\pi}{4} \vert k_r(r_{\rm ev})\vert \Delta r \; . \label{I parabolic}
}
Equation~(\ref{I parabolic}) shows that $\mathcal{I}$ is a proxy for the ratio of the thickness of the evanescent region to the characteristic length scale of decay of the wave amplitude in this region. In this toy model, the wave transmission coefficient $T^2$ in \eq{T parabolic} and thus the associated mode coupling factor $q$ computed through \eq{def q} depend only on this ratio (see \figurename{}~\ref{q vs a} for an illustration). 

First, in the limiting case of a very large ratio $\mathcal{I} \gg 1$, we have $T^2\rightarrow 0$ and $q\rightarrow 0$, so that the resonant cavities on both sides of the evanescent region are independent, as expected from \sectionname{}~\ref{resonance condition}. Then, in the intermediate case, the wave transmission increases as $\mathcal{I}$ decreases, so does the coupling factor. Finally, in the reverse limiting case of $\mathcal{I}=0$, $\Delta r =0$ and the incident wave energy is as much reflected as transmitted since \smash{$T^2 =1/2$} from \eq{T parabolic} and $R^2=1-T^2=1/2$ by energy conservation. This peculiar value of $1/2$ is mainly due to the symmetry with respect to $r=r_{\rm ev}$ of a parabolic evanescent barrier. Such a result can seem counterintuitive at first, since we can naively expect the wave transmission to be total when $\Delta r = 0$. It is actually not true since the incident wave still encounters a (double) turning point at $r=r_{\rm ev}$ and is still reflected. The associated mode coupling factor computed using \eq{def q} is equal to about \smash{$q~(\Delta r=0)\approx 0.17$}, which remains weak.

The previous example highlights two general points: reflection occurs as soon as the wave encounters a turning point, even if the thickness of the evanescent zone is null; a very thin evanescent zone is not necessarily associated with a strong coupling. 
It also shows that in general, the coupling factor is likely to depend on the wavenumber integral over the evanescent region and so on its radial extent. By analogy with this example, an evanescent zone is referred in the following to as thick when the wavenumber integral over its extent is much larger than unity, and it is referred to as thin when the wavenumber integral over its extent is equal to or smaller than unity.

%--------------------------------
\subsection{Asymptotic expression of the coupling factor}
%-------------------------------
\label{realistic structure}

The general expression of the coupling factor in stellar interiors were obtained by means of asymptotic analyses assuming only two limiting cases: either the case of a thick (i.e., $\mathcal{I}\gg1$) or a very thin (i.e., $\mathcal{I}\ll 1$) evanescent zone. In this paragraph, we introduce both formalisms that will form the theoretical basis of the study performed in this work.

%--------------------------------
\subsubsection{Hypothesis of a thick evanescent zone}
%-------------------------------
\label{weak coupling}

\cite{Shibahashi1979} first derived the resonance condition for mixed modes in \eq{resonance} under the hypothesis of a thick evanescent zone. In this case, the coupling factor is equal to
\algn{q &\approx \frac{T^2}{4} \label{q weak}}
and
\algn{
T^2 &= \exp\left( -2\mathcal{I}\right)\; , \label{T weak}
}
with $\mathcal{I}$ defined in \eq{k_r integral}. Equation~(\ref{q weak}) is consistent with the general definition of the coupling factor given in \eq{def q} in the limit of a low transmission coefficient (i.e., $T^2 \ll 1$). We also note that \eq{T parabolic} is consistent with \eq{T weak} in the limit of a thick evanescent zone (i.e., $\mathcal{I} \gg 1$). 

\cite{Shibahashi1979} originally derived \eq{q weak}{T weak}~using the Cowling approximation. We demonstrate in \appendixname{}~\ref{Cowling A} that the Cowling approximation is valid at leading order either where the short-wavelength WKB approximation is met or where the density contrast compared to the inner bulk is very high, that is, where $J\sim 1$. In the buoyancy and acoustic cavities, the WKB approximation is met. As a consequence, the error introduced by the Cowling approximation on the asymptotic period spacing or frequency large separation of mixed modes, which are leading-order terms in the asymptotic expansion, is thus negligible. In contrast, the error introduced on the asymptotic gravity offset of mixed modes, which corresponds to higher-order terms, is on the order of these parameters and thus cannot be neglected \citep[e.g.,][]{Takata2016a,Pincon2019}. Regarding the coupling factor, the WKB approximation is not valid close to the turning-points $r_1$ and $r_2$ associated with the evanescent region; the small deviation of $J$ from unity in this region turns out to be sufficient to modify the thickness of the evanescent region compared to the Cowling approximation, in particular in Type-a evanescent regions of subgiant and young red giant stars (see \figurename{}~\ref{propdiag} and \appendixname{}~\ref{Cowling B}). This explains the discrepancy observed between the values of the coupling factor computed with and without the Cowling approximation by \cite{Jiang2014} and \cite{Mosser2017b}. It appears thus necessary to consider the non-Cowling case when studying the coupling factor of dipolar modes. Actually, to transpose the result of \cite{Shibahashi1979} to the non-Cowling case, we have seen in \sectionname{}~\ref{dispersion} that we simply need to consider the modified Brunt-Väisälä and Lamb frequencies instead of their usual definitions inside the wavenumber integral $\mathcal{I}$.

Whether the Cowling approximation is used or not, \eq{q weak} shows that the value of the coupling factor cannot exceed 1/4 in the hypothesis of a thick evanescent region, in disagreement with observations \citep[e.g.,][]{Mosser2017b} and stellar models \citep[][]{Jiang2014,Hekker2018} for subgiant and red clump stars.

%-------------------------------
\subsubsection{Hypothesis of a very thin evanescent zone}
%-------------------------------
\label{strong coupling}

When the evanescent zone is very thin, the asymptotic development of the wave functions inside the evanescent zone, used by \cite{Shibahashi1979}, is not valid. Given the observational need to go beyond the hypothesis of a thick evanescent zone, \cite{Takata2016b} proposed a new asymptotic description of mixed modes in the reverse limiting case of a very thin evanescent zone. Solving the oscillation equations in this case while accounting for the perturbation of the gravitational potential, he found that the expression of the wave energy transmission coefficient reads 
\algn{T^2=\exp\left( -2 \mathcal{I} - \mathcal{G}_0\right) \; . \label{T strong}}
In this formulation, $T^2$ depends not only on the wavenumber integral $\mathcal{I}$, but also on a new term denoted with $\mathcal{G}_0$, which is related to the gradients of the equilibrium quantities inside the evanescent zone. The detailed analytical expression of this extra term is provided in \appendixname{}~\ref{G_0 general}. The coupling factor obtained through \eq{def q} can therefore vary between zero and unity, depending on the values of $\mathcal{I}$ and $\mathcal{G}_0$. Using typical stellar models, \cite{Mosser2017b} showed that this formulation can reproduce the global trend in the observed values of the coupling factor between the subgiant branch and the RGB. However, as said in \sectionname{}~\ref{introduction}, their interpretation in terms of internal structure called for a more detailed study. In this work, we propose to further examine the dependence of $q$ with the properties of the evanescent region within the asymptotic formulation in order to help interpret the coupling factor measurements.

\begin{table*}    
\caption{Summary of the assumptions considered in the analytical model for the subgiant and the red giant branches.}       
\centering          
\begin{tabular}{c|c | c| c |c }
\hline    
\hline
 Evanescent region & \multicolumn{2}{c|}{Type-a} & Transition regime& Type-b\\
 \hline       
   Evolutionary state &\multicolumn{2}{c|}{Subgiant and young red giant}&Red giant& Evolved red giant \\%\multicolumn{4}{c}{Method\#3}\\ 
 \hline
$\mathcal{N}$ and $\mathcal{S}_1$ & \multicolumn{2}{c|}{$\mathcal{N} (r) = \sigma \left( \dfrac{r_1}{r}\right)^{\beta},~\mathcal{S}_1(r) = \sigma\left( \dfrac{r_2}{r}\right)^{\beta} $} & Mixed& $\mathcal{N} (r) = 0,~\mathcal{S}_1(r) = \sigma\left( \dfrac{r_2}{r}\right)^{\beta} $ \\
\hline
$\mathcal{N}_{\rm b}$ and $\mathcal{S}_{\rm b}$ & \multicolumn{2}{c|}{$\mathcal{N}_{\rm b}~,~\mathcal{S}_{\rm b}~ <~ 2\pi\nu$} & $\mathcal{N}_{\rm b} < 2\pi\nu<\mathcal{S}_{\rm b}$& $2\pi\nu~<~\mathcal{N}_{\rm b}~,~ \mathcal{S}_{\rm b}$ \\
\hline
Location of $r_1~,~r_2$ & \multicolumn{2}{c|}{ $r_1~,~r_2~<~r_{\rm b}$} & $r_1<r_{\rm b}<r_2$ &$r_1 = r_{\rm b}<r_2$\\
\hline
$\beta$ and $J$ &\multicolumn{2}{c|}{$\beta = 3J/2$  }& Mixed  & $J\approx J(r_{\rm b})\approx 1$\\
\hline
Coupling formalism& \multicolumn{2}{c|}{Very thin (T16) or thick (S79)} & Thick (S79) & Thick (S79)\\
\hline    
\end{tabular}
\tablefoot{S79 and T16 denote the formalisms of \cite{Shibahashi1979} and \cite{Takata2016a}, respectively. The value of the (modified) Brunt-Väisälä and Lamb frequencies just below the base of the convective region are denoted with $\mathcal{N}_{\rm b}$ and $\mathcal{S}_{\rm b}$, respectively. ``Mixed'' means an evanescent region with a Type-a regime for $r_1<r<r_{\rm b}$ and a Type-b regime for $r_{\rm b}<r<r_2$, where $r$ is the radius inside the evanescent region (see \sectionname{}~\ref{transition} and \figurename{}~\ref{scheme}).}
\label{table model}    

\end{table*} 
%

%_________________________________
%
\section{Relation between $q$ and the properties of the evanescent zone}
%_________________________________
\label{simple model}

In this section, we investigate the relation that exists between the coupling factor of mixed modes and the evanescent region in an analytical way within the asymptotic framework. The model assumes that the profiles of the Brunt-Väisälä and Lamb frequencies follow the power-law behavior in \eq{power law N and S}{model B}~for Type-a and Type-b evanescent regions, respectively. This enables us to analytically express $q$ as a function of a few parameters characterizing the evanescent zone. The potential of $q$ to provide information on this region is subsequently analyzed. For sake of simplicity, we focus on stars on the subgiant and red giant branches; the case of red clump stars is addressed later in \sectionname{}~\ref{interpretation}.

\begin{figure}
\centering
\includegraphics[width=\hsize,trim= 0.8cm 0cm 0.8cm 1cm, clip]{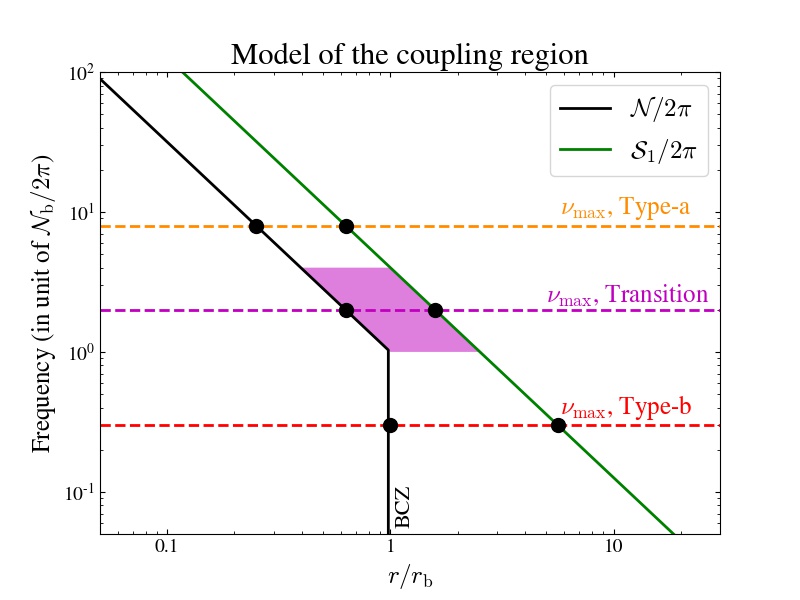} 
\caption{
Schematic view of the profiles of the (modified) Brunt-Väisälä and Lamb frequencies as a function of the radius normalized by the radius of the base of the convective zone, $r_{\rm b}$. The value of $\nu_{\rm max}$ in the case of Type-a and Type-b evanescent regions, as well as at the transition between both regimes, are indicated by the orange, magenta and red dashed lines, respectively. The black dots correspond to the turning points for the three cases. The magenta domain delimits all the possible locations of the evanescent regions in the transition regime.}
\label{scheme}
\end{figure}
%

%-----------------------------
\subsection{Analytical expression of $q$}
%---------------------------

To clarify the following computations, all the assumptions about the structure of the evanescent region are summarized in \tablename{}~\ref{table model}. A schematic view of the considered configuration is provided in \figurename{}~\ref{scheme}.

%---------------------------------
\subsubsection{Type-a evanescent zones}
%---------------------------------
\label{expression type a}

The values of $q$ measured by \cite{Mosser2017b} range between $0.15$ and $0.60$ for subgiant stars, and between about $0.2$ and $0.3$ for young red giant stars. The model SG in \figurename{}~\ref{propdiag} and the high observed values in the subgiant phase suggest to consider the asymptotic analysis of \cite{Takata2016a} at this stage. In contrast, the observed medium coupling values slightly higher or lower than 1/4, which is the upper limit predicted in the formalism of \cite{Shibahashi1979}, makes ambiguous the choice of the proper formalism to use, since the available asymptotic analyses consider only the limiting cases of thick and very thin evanescent zones.
Given these theoretical limitations, we thus consider both formalisms and compare them for Type-a evanescent regions.
In this framework, the coupling factor of mixed modes is deduced from \eq{def q}{T strong}~for very thin evanescent regions, and \eq{q weak}{T weak}~for thick evanescent regions. To obtain its expression, we thus need to compute the wavenumber integral $\mathcal{I}$ and the gradient-related term $\mathcal{G}_0$.

Firstly, the wavenumber integral in Type-a evanescent regions, denoted with $\mathcal{I}^r$, reads using \eqss{k_r}{Lamb}
\algn{
\mathcal{I}^r&=\int_{r_1}^{r_2} \vert k_r \vert {\rm d} r 
= \int_{r_1}^{r_2}  \frac{\sigma}{c}\left( 1 - \frac{\mathcal{N}^2}{\sigma^2}\right)^{1/2} \left(  \frac{\mathcal{S}_1^2}{\sigma^2}-1\right)^{1/2}  {\rm d} r \nonumber\\
&= \int_{r_1}^{r_2}  \sqrt{2} J  \frac{\sigma }{\mathcal{S}_1}\left( 1 - \frac{\mathcal{N}^2}{\mathcal{S}_1^2}\frac{\mathcal{S}_1^2}{\sigma^2}\right)^{1/2} \left(  \frac{\mathcal{S}_1^2}{\sigma^2}-1\right)^{1/2}  \frac{{\rm d} r}{r} \; .
\label{I^r 1}
}
For sake of convenience, we define in the following the parameter $\alpha$ such as
\begin{align}
\alpha\equiv \left(\frac{r_1}{r_2} \right)^\beta \; .
\label{alpha def}
\end{align}
The power-law assumption in \eq{power law N and S} shows that \smash{$\mathcal{N}(r)/\mathcal{S}_1(r)=\alpha$} in Type-a evanescent regions. This ratio is thus constant with respect to radius since $r_1$ and $r_2$ depend only on the oscillation frequency.
Using the change of variable $u=\sigma/\mathcal{S}_1$ in \eq{I^r 1}, the fact that $u(r_1)=\sigma/\mathcal{S}_1(r_1)=\mathcal{N}(r_1)/\mathcal{S}_1(r_1)=\alpha$ and \smash{$u(r_2)=\sigma/\mathcal{S}_1(r_2)=1$} in the bounds of the integral, as well as the equality $\beta=3J/2$, the wavenumber integral becomes \citep[e.g.,][]{Gradshteyn2007}
\algn{\mathcal{I}^r&=\int_{\alpha}^1  \frac{\sqrt{2}J}{\beta}\sqrt{1-u^2}\sqrt{u^2-\alpha^2} \frac{\dd u}{u^2}\nonumber\\
&=\frac{2 \sqrt{2}}{3}\left[ (\alpha^2+1) K(1-\alpha^2)-2 E(1-\alpha^2)\right] \; ,
\label{I^r para}
}
where $K$ are $E$ are the complete elliptic integrals of the first and second kind, respectively, in agreement with the result of \cite{Takata2016a}. 
In the case of a very thin evanescent region, $\alpha \rightarrow 1$ and $\mathcal{I}^r \rightarrow 0$. As $\alpha$ decreases, $\mathcal{I}^r$ increases since the evanescent region becomes thicker and thicker, as expected. When the evanescent zone becomes very thick, $\alpha \ll 1$ and we note for later purpose that the wavenumber integral in \eq{I^r para} tends to
\algn{\mathcal{I}^r&\approx \frac{2 \sqrt{2}}{3}\bigg[ -\ln \alpha+\ln 4 -2 +\circ (\alpha)\bigg] \; ,
\label{I^r para large}
}
where the term $\circ(\alpha)$ indicates negligible higher-order terms compared to $\alpha$.

\label{G_0}

\begin{figure}
\centering
\includegraphics[width=\hsize,trim= 0.8cm 0cm 0.8cm 1cm, clip]{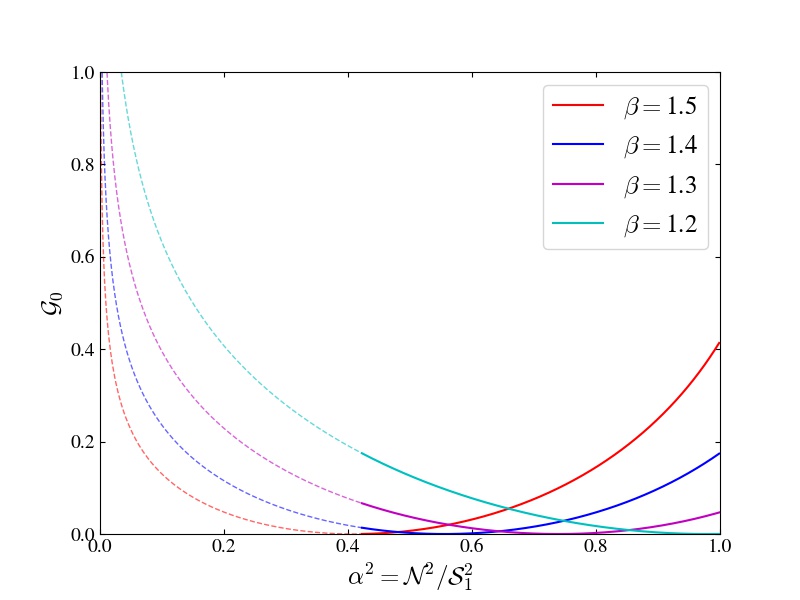} 
\caption{Gradient-related term $\mathcal{G}_0$ provided by \protect\eq{eq beta} as a function of $\alpha^2$ and for typical values of $\beta$. The dashed lines represent the extrapolation of this expression outside its derivation domain, that is, toward small values of $\alpha$ (i.e., for thick evanescent regions). The transition is arbitrarily assumed to occur when $\alpha^2 \lesssim 0.4$, i.e., when $\Delta r /2r_{\rm ev}\gtrsim 0.2$ according to \protect\eq{Dr vs alpha}.}
\label{G_0 fig}
\end{figure}

Secondly, the theoretical expression of the gradient-term in very thin evanescent regions, $\mathcal{G}_0$, which was obtained by \cite{Takata2016a}, is detailed in \appendixname{}~\ref{G_0 general}. Contrary to \cite{Mosser2017b}, we take the dependence of $\mathcal{G}_0$ on the thickness of the evanescent region into account. In the simple model represented by \eq{power law N and S}, this term is equal for Type-a evanescent regions to (see \appendixname{}~\ref{G_0 model})
\algn{
\mathcal{G}_0=\frac{3\pi}{4\sqrt{2} } \frac{\ln \alpha}{(\alpha-1)}  \left( \frac{1}{\ln \alpha} +\frac{\alpha}{1-\alpha}-\frac{1}{3}+\sqrt{ \frac{9}{4 \beta^2}-\frac{8\alpha^2}{9}}~\right)^2  \; .
\label{eq beta} 
}
In the limit case of a null thickness, that is, when $\alpha \rightarrow 1$, we retrieve the expression derived by \cite{Takata2016a} and considered in \cite{Mosser2017b} (see \eq{G_0 M17} for instance).
The predicted value of $\mathcal{G}_0$ computed with \eq{eq beta} is shown in \figurename{}~\ref{G_0 fig} as a function of the parameters $\alpha$ and $\beta$. In this figure, we see that in the case of very thin evanescent regions ($\alpha \sim 1$), the higher $\beta$, the higher $\mathcal{G}_0$. Indeed, the steeper the variation of the structure, the smaller the wave transmission through the evanescent barrier; therefore, according to \eq{T strong} with $\mathcal{I}\approx 0$, the higher $\mathcal{G}_0$, as expected. As $\alpha$ slightly decreases from unity, $\mathcal{G}_0 $ decreases, unlike the wavenumber integral. This decrease in $\mathcal{G}_0$ for $\alpha \lesssim 1$ can be understood as the fact that incident waves feel a smoother evanescent barrier as this latter becomes slightly larger (relatively to the local wavelength). 

Although it was derived in the hypothesis of a very thin evanescent region, it is interesting to discuss the extrapolation of the expression of $\mathcal{G}_0$ toward small values of $\alpha$ (i.e., for thick evanescent regions, which is outside of its derivation domain; see dashed lines in \figurename{}~\ref{G_0 fig}). We see that below a certain value of $\alpha$ where $\mathcal{G}_0$ vanishes, $\mathcal{G}_0$ diverges toward $+\infty$ as $\alpha$ tends to zero.
As shown by \eq{G_0 approx}, the extrapolation of $\mathcal{G}_0$ diverges as - $\ln \alpha$ when $\alpha\rightarrow 0$ over the considered range of values for $\beta$. According to \eq{I^r para large}, it is on the order of magnitude of the wavenumber integral given in \eq{I^r para large}.This prediction differs from the expectations of \cite{Takata2016a} who showed that, in general, the gradient-related term should vanish compared to the wavenumber integral when the evanescent region becomes large, because it represents a higher-order term in the asymptotic expansion. This example therefore emphasizes that the expression of $\mathcal{G}_0$ in \eq{def G_0}, from which is deduced \eq{eq beta}, is an approximation around $\alpha \approx 1$ of a more general expression of the gradient-related term that is valid whatever the thickness of the evanescent zone. In other words, the expression of the coupling factor computed using \eq{def G_0} does not converge toward the expression provided by \cite{Shibahashi1979} (i.e., $T^2$ in \eq{T strong} does not converge toward $T^2$ in \eq{T weak} as $\alpha\rightarrow 0$). This fact reinforces the ambiguity in the formalism to use to interpret medium values of the coupling factor, in the intermediate case between a thick and very thin evanescent region. This point is discussed in \sectionname{}~\ref{medium}.

%------------
\subsubsection{Type-b evanescent zones}
%-----------
\label{expression type b}

In typical evolved red giant stars with Type-b evanescent regions (i.e., for $\Delta \nu \lesssim 10~\mu$Hz), the shape of the evanescent region (e.g., model ER in \figurename{}~\ref{propdiag}) suggests to consider the weak coupling formalism of \cite{Shibahashi1979}; this choice is supported by the small values of $q$ close or lower than about $0.1$ inferred observationally by \cite{Mosser2017b} and in stellar models by \cite{Hekker2018}. 
We thus need to compute only the wavenumber integral in this case according to \eq{q weak}{T weak}.

Using \eq{model B} and the change of variable $u=\sigma/\mathcal{S}_1$, the wavenumber integral, denoted with $\mathcal{I}^c$ in this case, reads
\algn{
\mathcal{I}^c&
= \int_{r_{\rm b}}^{r_2}  \sqrt{2} J \left(1- \frac{\sigma^2}{\mathcal{S}_1^2}\right)^{1/2} \frac{ \dd r}{r} \nonumber \\
&=\int_{\alpha}^1\frac{ \sqrt{2} J}{\beta} \sqrt{1- u^2} \frac{ {\rm d} u}{u} \; ,
\label{I^c 1}
}
where we used at the boundaries of the integral the fact that $u(r_2)=1$ and $u(r_1)=\sigma/\mathcal{S}_1(r_1)=\alpha$ according to \eq{model B}{alpha def}.
Since $\alpha = \sigma / \mathcal{S}_1(r_{\rm b}) \ll1$ in general for Type-b evanescent regions (e.g., see model ER in \figurename{}~\ref{propdiag}), the integrand in \eq{I^c 1} is much larger near its lower bound than near its upper bound. Moreover, since the variation of $J$ over the evanescent region remains relatively small, the error made in the integral remains negligible if we assume that the value of $J$ is about equal to its value at the base of the convective zone. As a result, \eq{I^c 1} can be approximated by \citep[e.g.,][]{Gradshteyn2007}
\begin{align}
\mathcal{I}^c&\approx \frac{ \sqrt{2} J(r_{\rm b})}{\beta} \int_{\alpha}^1 \sqrt{1- u^2} \frac{ {\rm d} u}{u} \nonumber \\
&\approx \frac{ \sqrt{2}}{\beta}\left[ \ln \left( \sqrt{1-\alpha^2}+1\right)-\ln \alpha-\sqrt{1-\alpha^2}~\right] \; ,
\label{I^c 2}
\end{align}
where we also assumed $J(r_{\rm b}) \approx 1$ for evolved red giant stars, as shown for the model ER in \figurename{}~\ref{J graph}.
As expected, we find that $\mathcal{I}_{\rm c}$ increases as $\alpha$ decreases. When $\alpha \ll 1$ in the case of a thick Type-b evanescent zone, we note that the wavenumber integral in \eq{I^c 2} tends to
\begin{align}
\mathcal{I}^c&\approx  \frac{\sqrt{2}}{\beta}\bigg[ -\ln \alpha+\ln 2-1 +\circ (\alpha) \bigg]\; .
\label{I^c approx}
\end{align}
%

%------------------------------------------
\subsubsection{Degeneracy with the structural parameters}
%------------------------------------------
\label{dependence}

According to \eq{I^r para}{eq beta}{I^c 2}, the coupling factor of mixed modes depends on both parameters $\alpha$ and $\beta$. In the framework of the model, the knowledge of this couple of parameters is necessary and sufficient to fully characterize the evanescent region, which is basically represented by a certain thickness and a certain steepness of the profiles of $\mathcal{N}$ and $\mathcal{S}_1$.

On the one hand, the $\alpha$ parameter defined in \eq{alpha def} can actually be rewritten in the form of
\begin{align}
\alpha=\left(\frac{2-\Delta r/r_{\rm ev} }{2+\Delta r/r_{\rm ev} }\right)^\beta \; ,
\label{alpha vs Dr}
\end{align}
where $\Delta r=\vert r_2-r_1 \vert $ and $r_{\rm ev} =(r_1+r_2)/2$.
Equivalently, we can also write
\begin{align}
\frac{\Delta r}{r_{\rm ev} }= 2\left(\frac{1-\alpha^{1/\beta}}{1+\alpha^{1/\beta}}\right) \; .
\label{Dr vs alpha}
\end{align}
The $\alpha$ parameter is thus a proxy for the thickness of the evanescent region.

On the other hand, the $\beta$ parameter represents by definition the steepness of the profiles of $\mathcal{N}$ and $\mathcal{S}_1$ over the evanescent region. First, in Type-a evanescent zones, the high density contrast compared to the helium core implies \smash{$\beta \approx 3J/2$} over the region, which, included in \eq{B=f(J)} with $J$ regarded as locally uniform, leads to
\algn{
\beta_{\rm a} \approx -\frac{1}{2}\left(\deriv{\ln \rho}{\ln r}\right)_{r_{\rm ev}} \approx \frac{1}{2} \frac{r_{\rm ev}}{H_\rho(r_{\rm ev})}\; ,
}
where $H_\rho=-(\dd \ln \rho / \dd r)^{-1}$ is the local density scale height.
Second, in Type-b evanescent regions, which is located in the convective region, the equation of state is isentropic, so that $p\propto \rho^{\Gamma_1}$. As a result, $\beta$ can be expressed over the evanescent region as
\algn{
\beta_{\rm b}&\approx-\left(\deriv{\ln \mathcal{S}_1}{\ln r} \right)_{r_{\rm ev}}=-\frac{1}{2}\left(\deriv{\ln p}{\ln r}-\deriv{\ln \rho}{\ln r} \right)_{r_{\rm ev}}+1-\left(\deriv{\ln J}{\ln r}\right)_{r_{\rm ev}}\nonumber \\
&\approx \frac{1}{2}\left(\Gamma_1-1\right) \frac{r_{\rm ev}}{H_\rho(r_{\rm ev})}+1\; ,
}
where \smash{$(\dd \ln J/\dd \ln r )_{r_{\rm ev}}$} is neglected at first approximation. Therefore, $\beta$ turns out to be directly related to the local density scale height, $H_{\rho}$.

To conclude, the model shows us that the coupling factor of mixed modes depends on the thickness of the evanescent zone represented by $\Delta r/r_{\rm ev}$, and on the local density scale height represented by $\beta$. The value of $q$ is therefore in general degenerate with respect to these two internal properties whatever the type of the evanescent region.

%------------------------
\subsection{Variations of $q$ with respect to the structure}
%-----------------------

As shown in the previous section, the link between $q$ and the structural parameters $\beta$ and $\Delta r/r_{\rm ev}$ depends on the type of the evanescent region. Both types of evanescent region are thus taken into account and compared in the present section. Given the ambiguity in the applicability of the weak or the strong coupling formalisms regarding the thickness of the evanescent region (see \sectionname{}~\ref{expression type a}), we compute and compare both expressions of $q$ as provided by \eq{T weak}{T strong}~for a large range of values of $\Delta r/r_{\rm ev}$. Assumptions on the applicability domain of each formalism will be made in a second step for the interpretation of the observations (see \sectionname{}~\ref{interpretation}).

%----------------------
\subsubsection{Type-a evanescent regions}
%-----------------------
\label{Type-a}

\begin{figure}
\centering
\includegraphics[width=\hsize,trim= 0.8cm 0cm 0.8cm 1cm, clip]{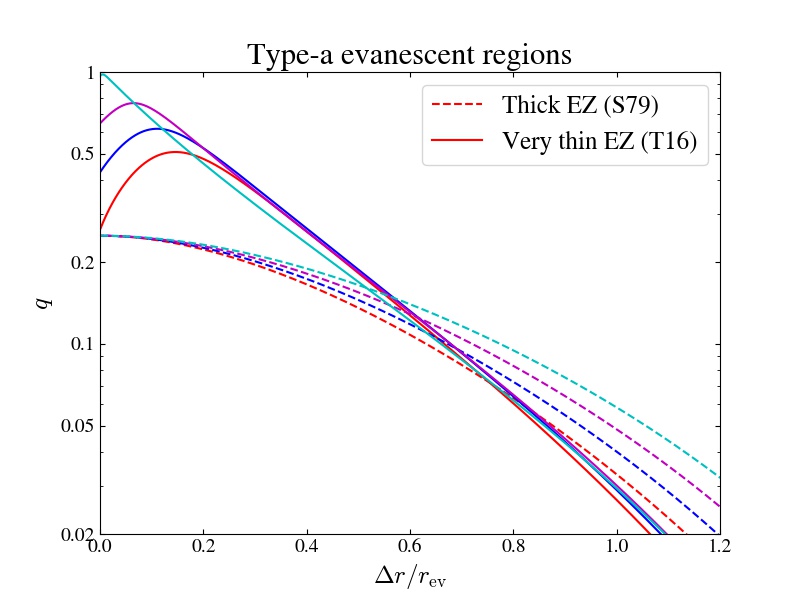} 
\caption{
Coupling factor as a function of the normalized radial extent of the evanescent region in Type-a evanescent regions, as expected from the model. The cyan, blue, magenta and red colors represent the result for a typical range of values of the $\beta$ parameter, i.e., $\beta=1.2,~1.3,~1.4$, and $1.5$, respectively. The dashed lines correspond to the predictions obtained with the formalism of \cite{Shibahashi1979} according to \protect\eq{q weak}{T weak}{I^r para}. The solid lines correspond to the predictions computed in the formalism of \cite{Takata2016a} according to \protect\eq{def q}{T strong}{I^r para}{eq beta}.
}
\label{q vs Dr a}
\end{figure}

The prediction of the model in Type-a evanescent regions is plotted in \figurename{}~\ref{q vs Dr a}. In this figure, the coupling factor was computed for typical values of $\beta$ using either \eq{q weak}{T weak}{I^r para}~in the formalism of \cite{Shibahashi1979}, or \eq{def q}{T strong}{I^r para}{eq beta}~in the formalism of \cite{Takata2016a}, as well as \eq{Dr vs alpha}. The values of $q$ were taken in the range between $0.02$ and $1$, which encapsulates the observed range of values \citep[e.g.,][]{Mosser2017b}.

In the formalism of \cite{Shibahashi1979}, \figurename{}~\ref{q vs Dr a} shows that the lower the value of $\Delta r /r_{\rm ev}$ or $\beta$, the higher the value of $q$ . Indeed, using for instance the approximation in \eq{I^r para large} in the limit of a very thick evanescent region, the coupling factor can be rewritten according to \eq{q weak}{T weak}~such as
\algn{
q_{\rm a} \approx 0.8 \alpha ^{4\sqrt{2}/3} \; .
\label{q approx thick}
}
Since $\alpha$ decreases as $\Delta r/ r_{\rm ev}$ or $\beta$ increases according to \eq{alpha vs Dr}, \eq{q approx thick} confirms that $q$ must also decrease. This trend originates from the fact that the wave transmission factor through the evanescent region is lower either when this region is larger or when the length scale of decay of the wave amplitude is smaller (i.e., when $\beta$ is higher). We also see in \figurename{}~\ref{q vs Dr a} that at a given value of $\Delta r/r_{\rm ev}$, the influence of $\beta$ is negligible for $q\sim 0.20$, but significant for $q\sim 0.05$ (with a change in $q$ of about 60\% between $\beta=1.5$ and $\beta=1.2$). The dependence of $q$ on $\beta$, which comes from the $\alpha$ parameter in \eq{q approx thick}, thus cannot be neglected.

In contrast, in the formalism of \cite{Takata2016a}, \figurename{}~\ref{q vs Dr a} shows that $q$ depends on $\Delta r/r_{\rm ev}$ and $\beta$ following two different regimes. At low values of $\Delta r /r_{\rm ev}$, that is, between about $0$ and $0.15$, the predicted values of $q$ are high, comprised between about 0.25 and 1. The lower $\beta$, the higher $q$, but the lower $\Delta r /r_{\rm ev}$, the lower $q$. This difference of trend compared to the formalism of \cite{Shibahashi1979} results from the predominant effect of the gradient-related term $\mathcal{G}_0$ in this case (see \figurename{}~\ref{G_0 fig}). This term also results in a more important sensitivity of $q$ to the parameter $\beta$ in this regime. Indeed, at a given $\Delta r/r_{\rm ev}$, a change in $\beta$ between $1.2$ and $1.5$ results in a large change of $q$ by a factor of between two and four. At values of $\Delta r /r_{\rm ev}\gtrsim 0.15$, \figurename{}~\ref{q vs Dr a} shows that the higher $\Delta r /r_{\rm ev}$, the lower $q$, but no substantial dependence on $\beta$ can be depicted. This results again to the gradient-related term $\mathcal{G}_0$ whose value is large enough to counterbalance the effect of the wavenumber integral (see second paragraph in \sectionname{}~\ref{expression type a}).

In summary, the behavior of $q$ with respect to the properties of the evanescent region is substantially different depending on the formalism we consider. In the formalism of \cite{Shibahashi1979}, $q$ decreases as $\Delta r /r_{\rm ev}$ or $\beta$ increases, with a substantial dependence on $\beta$. In the formalism of \cite{Takata2016a}, the effect of the gradient-related term $\mathcal{G}_0$ is important and drastically modifies how $q$ varies with respect to $\Delta r /r_{\rm ev}$ and $\beta$. In particular, for $\Delta r /r_{\rm ev}\lesssim 0.15$, $q$ turns out to increase as $\Delta r /r_{\rm ev}$ increases and to be very sensitive to $\beta$ over the typical considered range of values.

%---------------------
\subsubsection{Type-b evanescent regions}
%---------------------
\label{Type-b}

\begin{figure}
\centering
\includegraphics[width=\hsize,trim= 0.8cm 0cm 0.8cm 1cm, clip]{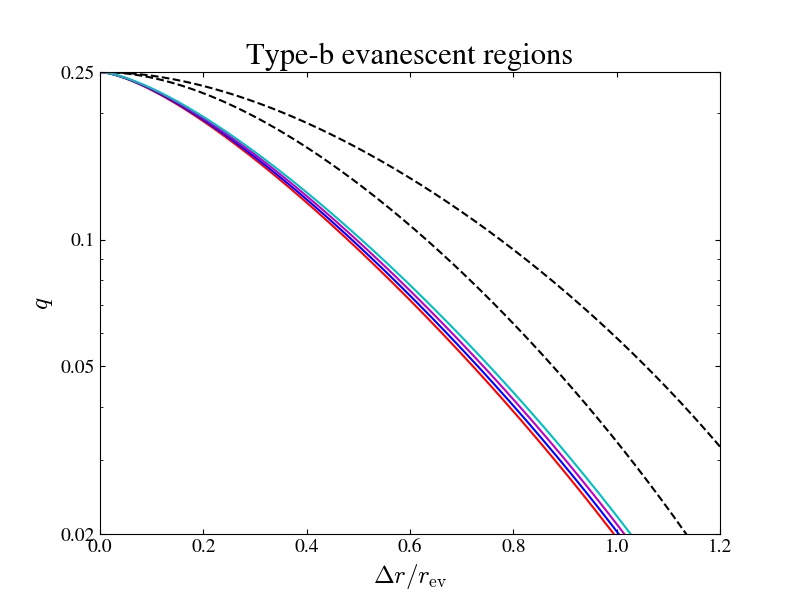} 
\caption{
Coupling factor as a function of the normalized radial extent of the evanescent region in Type-b evanescent regions according to Eqs.~(\ref{q weak}),~(\ref{T weak})~and~(\ref{I^c 2}). The cyan, blue, magenta and red colors have the same meaning as in \figurename{}~\ref{q vs Dr a}. For comparison, the domain between both dashed lines represents the predictions obtained for Type-a evanescent regions in the formalism of \cite{Shibahashi1979} such as plotted in \figurename{}~\ref{q vs Dr a}.
}
\label{q vs Dr b}
\end{figure}
The predictions of the model in Type-b evanescent regions are shown in \figurename{}~\ref{q vs Dr b}. In this figure, $q$ was computed for typical values of $\beta$ using \eq{q weak}{T weak}{I^c 2}~in the formalism of \cite{Shibahashi1979} as well as \eq{Dr vs alpha}. The values of $q$ were chosen between $0.02$ and its upper limit equal to $0.25$.

As seen in \figurename{}~\ref{q vs Dr b}, the higher $\Delta r /r_{\rm ev}$ or $\beta$, the lower $q$ for the same reasons as in Type-a evanescent regions in the thick limiting case. However, \figurename{}~\ref{q vs Dr b} first shows that at a given value of $\Delta r / r_{\rm ev}$, the value of $q$ obtained in Type-b evanescent zones is about twice lower than in Type-a evanescent regions, and much less sensitive to $\beta$. To explain this difference, it is convenient to use \eq{I^c approx} in the limit of a very thick evanescent region; the coupling factor can thus be expressed according to \eq{q weak}{T weak}{Dr vs alpha}~such as
\algn{
q_{\rm b} \approx \frac{2.38^{2\sqrt{2}/\beta}}{4} \alpha^{2\sqrt{2}/\beta} = \frac{2.38^{2\sqrt{2}/\beta}}{4} \left(\frac{2-\Delta r/r_{\rm ev} }{2+\Delta r/r_{\rm ev} }\right)^{2\sqrt{2}} \; .
\label{q b}
}
The first multiplicative factor in \eq{q b} only slightly varies between 0.51 and 0.45 (i.e., lower than 10\%) as $\beta$ changes from 1.2 and 1.5. Therefore, the comparison between \eq{q approx thick}{q b}~confirms the difference depicted in \figurename{}~\ref{q vs Dr b} between Type-a and Type-b evanescent regions. This is the consequence of two facts. First, $N^2\approx 0$ in convective regions so that the wavenumber integral over the evanescent region is higher and thus the wave transmission factor is lower; second, the linear relation between $\beta$ and $J$ is not valid in this case.

%------------------------
\subsection{Transition from Type-a to Type-b evanescent regions}
%-----------------------
\label{transition}

As seen in the previous section, the value of $q$ is about twice lower in a Type-a than in a Type-b evanescent region for the same values of $\beta$ and $\Delta r /r_{\rm ev}$. At first sight, the transition from a Type-a to a Type-b evanescent region during evolution could thus be expected to be associated with a rapid variation in the value of $q$ measured around $\nu_{\rm max}$. Nevertheless, the transition actually does not occur at a given instant (i.e., a given value of $\nu_{\rm max}$), but is composed of an intermediate phase where the evanescent region progressively changes from a Type-a to a Type-b regime. Indeed, as stars evolve, the Type-a regime ends as soon as $r_2$ becomes higher than $r_{\rm b}$, while $r_1$ still remains lower than $r_{\rm b}$. This happens when $\mathcal{N}_{\rm b}\lesssim2\pi \nu \lesssim \mathcal{S}_{\rm b}$, where $\mathcal{S}_{\rm b}=\mathcal{S}_1(r_{\rm b})$ is the value of the Lamb frequency at the base of the convective region. During this transition phase, a Type-a configuration is met for $r_1 <r<r_{\rm b}$ whereas a Type-b configuration is met for $r_{\rm b} < r < r_2$. The Type-b regime then starts when $r_1\approx r_{\rm b}$, that is, when $2\pi \nu \lesssim \mathcal{N}_{\rm b}$ (see \tablename{}~\ref{table model} and \figurename{}~\ref{scheme}).

To estimate the effect of this progressive transition on the coupling factor in the framework of the analytical model, the wavenumber integral over the evanescent region when \smash{$r_1<r_{\rm b}<r_2$}, denoted with $\mathcal{I}^{\rm t}$, has to be computed. It can be easily obtained by combining \eq{I^r para}{I^c 2}~while adapting the upper bound of \eq{I^r para}, that is, such as
\algn{
\mathcal{I}^{\rm t}\approx \frac{2\sqrt{2}}{3} \int_{\alpha_{\rm b}}^{u_{\rm b}}  \sqrt{1-u^2}\sqrt{u^2-\alpha_{\rm b}^2} \frac{\dd u}{u^2} +\frac{\sqrt{2}}{\beta}\int_{u_{\rm b}}^1 \sqrt{1- u^2} \frac{ {\rm d} u}{u}\; ,
\label{I^t 1}
}
with
\algn{
\alpha_{\rm b}=\frac{\mathcal{N}_{\rm b}}{\mathcal{S}_{\rm b}}~~~~~~{\rm and}~~~~~~\; u_{\rm b} =\frac{\sigma}{\mathcal{S}_{\rm b}}=\alpha_{\rm b} \frac{\sigma}{\mathcal{N}_{\rm b}}\; ,
\label{u_b def}
}
where we used in the lower bound of the first integral the fact that $u(r_1)=\sigma/\mathcal{S}_1(r_1)=\mathcal{N}(r_1)/\mathcal{S}_1(r_1)=\mathcal{N}_{\rm b}/\mathcal{S}_{\rm b}$ since this ratio is independent of radius for $r\le r_{\rm b}$ (see \sectionname{}~\ref{expression type a}).
To go further, the integral in \eq{I^t 1} can be expressed as a function of special functions to finally obtain \citep[e.g.,][]{Gradshteyn2007}
\algn{
&\mathcal{I}^{\rm t}\approx\frac{\sqrt{2}}{\beta} \left[~\ln \left( \sqrt{1-u_{\rm b}^2}+1\right)-\ln u_{\rm b}-\sqrt{1-u_{\rm b}^2} ~~\right]\nonumber\\
&-\frac{2\sqrt{2}}{3}\left[\sqrt{(1-u_{\rm b}^2)\left(1-\frac{\alpha_{\rm b}^2}{u_{\rm b}^2}\right)}-2E(\gamma_{\rm b}~\vert~1-\alpha_{\rm b}^2)+2E(1-\alpha_{\rm b}^2)~ \right]\nonumber\\
&-\frac{2\sqrt{2}}{3}(1+\alpha_{\rm b}^2)\left[F(\gamma_{\rm b}~\vert~1-\alpha_{\rm b}^2)-K(1-\alpha_{\rm b}^2)~\right] \; ,
\label{I_t}
}
where $K(x)$ are $E(x)$ and the complete elliptic integrals of the first and second kind, $F(z~\vert~x)$ and $E(z~\vert~x)$ are the incomplete elliptic integrals of the first and second kind of parameter $k^2=x$, and 
\algn{
\gamma_{\rm b}=\arcsin\left(\sqrt{\frac{1-u_{\rm b}^2}{1-\alpha_{\rm b}^2}} \right) \; .
}
Additional simplifying assumptions can be made to go a step further. Since the considered red giant stars close to the transition are evolved enough, we first assume that the evanescent region is thick and use the weak coupling formalism of \cite{Shibahashi1979}. Second, we assume that $\beta$ is nearly uniform and equal to 1.5 on both sides of $r_{\rm b}$. This is a reasonable assumption since, on the one hand, the density contrast in the vicinity of the base of the convective region compared to the helium core is so high that $J\approx 1$ in the radiative region, and thus $\beta \approx 3J/2\approx 1.5$ here. On the other hand, the sensitivity of $q$ to $\beta$ in the convective region is small enough to be neglected at first approximation (see \sectionname{}~\ref{Type-b}), so that choosing $\beta \approx 1.5$ has only a small impact on the final result.

\begin{figure}
\centering
\includegraphics[width=\hsize,trim= 0.8cm 0cm 0.8cm 1cm, clip]{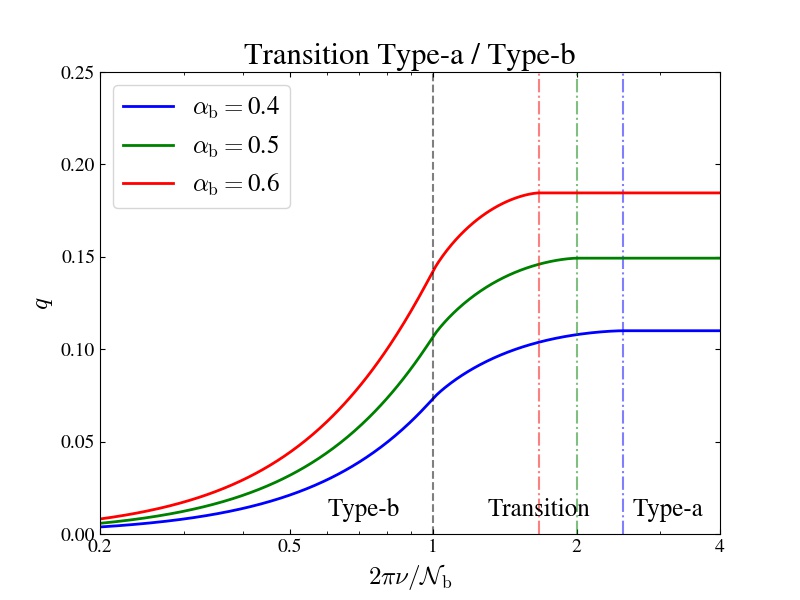} 
\caption{Evolution of the coupling factor as a function the ratio $2\pi \nu/\mathcal{N}_{\rm b}$ (solid lines), such as predicted in \sectionname{}~\ref{transition} and for different values of $\alpha_{\rm b}=\mathcal{S}_{\rm b} /\mathcal{N}_{\rm b}$, where $\mathcal{N}_{\rm b}$ and $\mathcal{S}_{\rm b}$ are the values of the modified Brunt-Väisälä and Lamb frequencies just below the base of the convective region. When considering $\nu\approx\nu_{\rm max}$, stellar evolution goes from the right to the left. The black dashed line indicates the beginning of the Type-b regime. The dash-dotted lines delimit the end of the Type-a regime for the considered values of $\alpha_{\rm b}$, with the corresponding colors.}
\label{q_Nb}
\end{figure}

Within these considerations, the value of the coupling factor can be computed as a function of $2\pi \nu/\mathcal{N}_{\rm b}$ using \eq{q weak}{T weak}, and considering for the wavenumber integral:
\begin{itemize}
\item $\mathcal{I}^r$ in \eq{I^r para} if $\alpha_{\rm b}^{-1} < 2\pi\nu /\mathcal{N}_{\rm b}$;
\item $\mathcal{I}^{\rm t}$ in \eq{I_t} if $1<2\pi \nu/\mathcal{N}_{\rm b} < \alpha_{\rm b}^{-1}$;
\item and $\mathcal{I}^c$ in \eq{I^c 2} if $ 2\pi\nu/\mathcal{N}_{\rm b}<1$. 
\end{itemize}
The range of values for $\alpha_{\rm b}$ is chosen between about 0.4 and 0.6, as required by the model to reproduce typical values of $q$ observed between about $0.1$ and $0.2$ in stars with $10~\mu$Hz~$\lesssim \Delta \nu \lesssim 18 ~\mu$Hz \citep{Mosser2017b,Hekker2018}, which correspond to stars before the expected transition to the Type-b regime (see \sectionname{}~\ref{Prop diag}). The result is plotted in \figurename{}~\ref{q_Nb}. From an observational point of view, the ratio $2\pi \nu /\mathcal{N}_{\rm b}$ for $\nu \sim \nu_{\rm max}$ generally decreases during the evolution on the RGB. Indeed, the decrease in $\nu_{\rm max}$ induced by the envelope expansion is predominant compared to the variation in $\mathcal{N}_{\rm b}$ that is mainly induced by variations in the location of the base of the convective region, as shown by the rough estimate $\mathcal{N}_{\rm b}^2 \sim G M_r(r_{\rm b})/r_{\rm b}^3$. The value of this ratio is considered between 0.2 and 4, which is representative of the ratio $2\pi \nu_{\rm max}/\mathcal{N}_{\rm b}$ in typical models of red giant stars close to the transition and just before the luminosity bump \citep[][see their Appendix~B for details]{Pincon2019}.

We see in \figurename{}~\ref{q_Nb} that $q$ decreases as $2\pi \nu /\mathcal{N}_{\rm b}$ or $\alpha_{\rm b}$ decreases whatever the regime considered. Indeed, first, in the Type-b regime, we can express the parameter $\alpha$ in \eq{alpha def} as $\alpha=2\pi \nu  \alpha_{\rm b}/ \mathcal{N}_{\rm b} $ according to \eq{model B}~with $r_1= r_{\rm b}$. The last equality shows that the decrease in $q$ as $2\pi \nu /\mathcal{N}_{\rm b}$ or $\alpha_{\rm b}$ decreases actually results from a decrease in $\alpha$, or equivalently, from an increase in $\Delta r /r_{\rm ev}$, as already shown in \sectionname{}~\ref{Type-b}. Second, in the Type-a and transition regimes, we have $\alpha=\alpha_{\rm b}$ since $\alpha$ does not depend on radius (i.e., when $r<r_{\rm b}$). Therefore, at a fixed value of $2\pi \nu/\mathcal{N}_{\rm b}$, the decrease in $q$ as $\alpha_{\rm b}$ (i.e., $\alpha$) decreases results also from an increase in $\Delta r/r_{\rm ev}$ (since $\beta$ remains close to 1.5). However, in the transition regime, an increase in $\Delta r/r_{\rm ev}$ is not the only possible reason for the decrease in $q$ observed in \figurename{}~\ref{q_Nb}. Indeed, at a fixed value of $\alpha_{\rm b}$, $\alpha$ and $\Delta r/r_{\rm ev}$ are also fixed, but $q$ still decreases as $2\pi \nu / \mathcal{N}_{\rm b}$ decreases. In fact, this decrease in $q$ results, according to \eq{I^t 1}{u_b def}, from the progressive migration of the evanescent region from the radiative to the convective zones, because of the simultaneous progressive increase in the contribution of the integral between $r_{\rm b}$ and $r_2$ to the total wavenumber integral (and since the wavenumber integral over any given region is always higher when $\mathcal{N}^2\approx0$).

The analytical predictions can now be used to address the variation of the observed coupling factor $q(\nu_{\rm max})$ for $\nu\sim\nu_{\rm max}$ during the evolution on the RGB. During this phase, the parameter $\alpha_{\rm b}=\mathcal{N}_{\rm b}/\mathcal{S}_{\rm b}$ can be assumed to vary much more slowly than $2\pi \nu_{\rm max} /\mathcal{N}_{\rm b}$, because both $\mathcal{N}^2$ and $\mathcal{S}_1^2$ similarly evolve as $g/r$ in the radiative region whereas $\nu_{\rm max}$ decreases more rapidly than $\mathcal{N}_{\rm b}$ varies (see \sectionname{}~\ref{Prop diag}). Therefore, as $2\pi \nu_{\rm max} / \mathcal{N}_{\rm b}$ decreases (from large values to unity) at a quasi constant value of $\alpha_{\rm b}$, the model first predicts that $q(\nu_{\rm max})$ starts smoothly decreasing at the beginning of the transition phase because of the progressive migration of the evanescent region from the radiative to the convective zones. Second, when the evanescent region is entirely located in the convective region for $2\pi \nu_{\rm max} / \mathcal{N}_{\rm b}\lesssim 1$, $q(\nu_{\rm max})$ keeps progressively decreasing along with the evolution because $\Delta r /r_{\rm ev}(\nu_{\rm max})$ increases as $\nu_{\rm max}$ decreases.

Therefore, we conclude that no sharp variation in $q(\nu_{\rm max})$ is expected at the transition from a Type-a to a Type-b evanescent region. Instead, our simplified estimate predicts that $q(\nu_{\rm max})$ smoothly and progressively decreases at this stage, mainly because of the progressive migration of the evanescent region from the radiative to the convective regions (provided that $\alpha_{\rm b}$ does not vary too rapidly). This is in contrast with the gravity offset of mixed modes, denoted with $\varepsilon_{\rm g}$, which is another seismic parameter associated with mixed modes appearing inside the term $\Theta_{\rm g}$ of \eq{resonance}. This parameter was shown by \cite{Pincon2019} to exhibit a clear drop at the transition to the Type-b regime because of the kink in the Brunt-Väisälä frequency at the base of convective region. The authors did not notice any obvious simultaneous signature in the observed values of $q(\nu_{\rm max})$ for such evolved stars; the analytical model developed in the present work confirms the absence of such a sharp variation in the global evolution of $q(\nu_{\rm max})$.

%

%------------------------------
\subsection{Lessons learned in a nutshell}
%------------------------------

To summarize this first part of the study, the structure of the evanescent coupling region changes during evolution. In subgiant, young red giant and red clump stars, the evanescent region is located in the radiative region, where $\mathcal{N}$ and $\mathcal{S}_1$ vary at first approximation as power laws of radius with the same exponent $\beta$ (Type-a regime). In evolved RGB stars, with typically $\Delta \nu \lesssim 10 ~\mu$Hz, the evanescent region is located in the lower part of the convective region, where $\mathcal{N}\approx 0$ (Type-b regime).

In both cases, the analytical model developed shows that the value of $q$ is degenerate with respect to $\beta$ and $\Delta r /r_{\rm ev}$. In particular, the model highlights the following points:
\begin{itemize}
\item In the Type-a regime, the behavior of $q$ with both parameters significantly changes depending on the thickness of the evanescent region. In thick evanescent regions, $q$ monotonically decreases as $\beta$ or $\Delta r /r_{\rm ev}$ increases. In very thin evanescent regions, $q$ turns out to be very sensitive to $\beta$ (for a typical range of values), in a much more important way than in thick evanescent regions; moreover, $q$ in contrast increases as $\Delta r /r_{\rm ev}$ increases for $\Delta r /r_{\rm ev}\lesssim 0.15$. This drastic change of trend originates from the effect of the gradient in the equilibrium structure on the wave transmission through the evanescent region, represented by $\mathcal{G}_0$ in \eq{T strong}.
\item In Type-b evanescent regions, $q$ also decreases as $\beta$ or $\Delta r /r_{\rm ev}$ increases. The value of $q$ is nevertheless about twice lower than in Type-a regions for the same couple of parameters $\beta$ and $\Delta r /r_{\rm ev}$. Moreover, its sensitivity to $\beta$ is very small; the value of $q$ measured around $\nu_{\rm max}$ is thus expected to decrease during the evolution, because of the progressive increase in $\Delta r /r_{\rm ev} (\nu_{\rm max})$ as $\nu_{\rm max}$ decreases.
\item Unlike the gravity offset of mixed modes, $q(\nu_{\rm max})$ is not expected to exhibit a sharp variation at the transition from the Type-a to the Type-b regimes on the RGB, which occurs just before the luminosity bump.
\end{itemize}

Finally, we emphasize that the trend of $q$ with $\Delta r /r_{\rm ev}$ predicted by the analytical model in evolved red giant stars with Type-b evanescent regions is comparable to that found in stellar models by \citet[][see their Fig.~7]{Hekker2018}, at least for values of $q$ lower than about $0.15$. However, the result of their fit is questionable for the red clump stars they considered in the strong coupling case since the computation of \cite{Hekker2018} does not account for the gradient-term $\mathcal{G}_0$, which turns out to play an important role in Type-a evanescent regions, as demonstrated in \sectionname{}~\ref{Type-a}.

%_______________________________________
%
\section{Clues on the probing potential of $q$}
%_______________________________________
\label{interpretation}

In this section, we investigate the ability of the coupling factor to bring us constraints on stellar interiors. We give a qualitative interpretation of the variations in $q$ observed by \cite{Mosser2017b} in the light of the analytical model developed in the previous section. The discussion is illustrated by considering typical values of $q$ and their error bars in subgiant (SG), young red giant (YR) and evolved red giant stars (ER). The case of red clump stars (RC) is also considered. The confrontation of the model to observations implicitly assumes that $q$ is taken at the frequency $\nu_{\rm max}$ for a given star. The possible dependence of $q$ with the frequency over the observed oscillation spectrum is addressed in a subsequent step. The present study represents a preparatory stage toward a more quantitative investigation using stellar models that will be presented in the paper II of the series.

%------------------
\subsection{Qualitative interpretation of the current measurements}
%------------------

%------------------
\subsubsection{Observations}
%-----------------
%
\begin{table*}
\caption{Typical values of $q$ and theoretical framework used to interpret the observations with respect to the evolutionary stage.}   
\centering  
\begin{tabular}{cccccc}     % 7 columns 
\hline\hline       
                      % To combine 4 columns into a single one 
&$\Delta \nu$ ($\mu$Hz)&$\nu_{\rm max}$ ($\mu$Hz)&$q$&\multicolumn{2}{c}{Evanescent region model}\\\cline{5-6}
\multicolumn{4}{c}{}&~~~~~~Type~~~~~~&Thickness\\ 
\hline                   
Subgiant (SG)&$\gtrsim30$&$\gtrsim500$&$\sim0.15-0.65$&a&thick to very thin\\
               
Young red giant (YR)&$\sim18-30$& $\sim 250-500$&$\sim0.18-0.30$&a & very thin to thick\\

Transition&$\sim10-18$&$\sim 110-250$&$\sim0.13-0.18$&mixed& thick\\

Evolved red giant (ER)&$\lesssim5-10$&$\lesssim 50-110$&$\lesssim 0.13 $&b&thick\\
 
Red clump&$\sim 3 -8~\mu$Hz&$\sim20-100$&$\sim 0.3$&a&very thin\\
\hline
\end{tabular}
\label{table model 2}
\tablefoot{The values of $\nu_{\rm max}$ is estimated through the usual $\Delta \nu-\nu_{\rm max}$ scaling law for RGB stars \citep[e.g.,][]{Mosser2012b}.}
\end{table*}

The large-scale measurement of $q$ as a function of $\Delta \nu$ (or $\nu_{\rm max}$) by \citet[][see their Figs. 5 and 6]{Mosser2017b} shows that stars exhibit low values of $q$ around $0.15$ at the very beginning of the subgiant branch before rapidly increasing toward a peak value of about $0.5$. During this peak phase on the subgiant branch, the spread in the measurements is large, on the order of about $0.15$.

In the following stages for typically $18~\mu$Hz$\lesssim \Delta \nu \lesssim 30~\mu$Hz, the observed value of $q$ decreases from about $0.30$ to $0.18$. During the subsequent evolution on the RGB, the coupling factor continues to decrease slowly in average, with a spread in the measurements lower than about $0.03$, that is, much smaller than in subgiant stars. The most luminous red giant stars of the sample with $\Delta \nu \sim 5~\mu$Hz have values of $q$ around $0.10$, with an observed minimum value close to $0.05$.

Regarding red clump stars, the observed values of $q$ remain around a high mean value of about $ 0.3$. This suggests that these stars are in the regime of very thin (Type-a) evanescent regions. The spread in the measurements is low and close to about $0.02$ around the mean value\footnote{This result was obtained only for stars with high-quality fits, which still represent a significant statistical sample.}.

%---------------------------------
\subsubsection{Coupling regimes vs evolutionary stage}
%---------------------------------

The analytical expressions of $q$ depend both on the type and on the hypothesis about the thickness of the evanescent region (i.e., thick or very thin). To interpret observations of $q$ in the framework of the model along with evolution, we thus have to make additional assumptions about which type and coupling formalism to consider depending on the evolutionary state and the value of $q$.

First, we estimate through standard stellar models that evolved red giant stars with Type-b evanescent regions have $\Delta \nu \lesssim 5$ and $10~\mu$Hz for stellar masses between $2M_\odot$ and $1M_\odot$, respectively (see \sectionname{}~\ref{Prop diag}). \cite{Mosser2017b} observationally determined that $q$ on the RGB  follows the fit law $q\approx q_{\rm M} (\Delta \nu/10)^{0.1}$, in  which $q_{\rm M}\in [0.125,0.145]$ in the considered mass range and $\Delta \nu$ is expressed in $\mu$Hz. According to this relation, we thus consider that (thick) Type-b evanescent regions are met when $q\lesssim 0.13$ in RGB stars.

Just above $q \approx 0.13$ on the RGB, stars are in the transition regime. In this regime, $q$ depends not only on $\Delta r /r_{\rm ev}$ and $\beta$, but also on the ratio $2\pi \nu_{\rm max}/\mathcal{N}_{\rm b}$, which represents the influence of the migration of the evanescent region from the radiative to the convective region. Using the theoretical picture presented in \sectionname{}~\ref{transition} (assuming thick evanescent regions), we find that the model requires $\alpha_{\rm b} \approx 0.55$ to explain $q\approx 0.13$ at the end of the transition phase (i.e., when $2\pi \nu_{\rm max}/\mathcal{N}_{\rm b} \approx 1$, see \figurename{}~\ref{q_Nb}). Assuming that $\alpha_{\rm b}=\mathcal{N}_{\rm b}/\mathcal{S}_{\rm b}$ is fixed at $0.55$ during this phase because of the similar evolution of $\mathcal{N}^2$ and $\mathcal{S}_1^2$ as $g/r$ in the radiative zone, the model predicts $q\approx0.18$ at the beginning of the transition regime. We thus fix that the Type-a regime is met for $q\gtrsim 0.18$ on the RGB.

In young red giant stars, corresponding to RGB stars with $q\gtrsim 0.18$ following the previous estimate, as well as in subgiant stars with typically $q \approx 0.15-0.65$, thick and very thin Type-a evanescent regions are considered and compared owing to the ambiguity in the coupling formalism to use. For red clump stars, we consider very thin (Type-a) evanescent regions since this is the only current formalism able to reproduce the high observed values close to $0.3$.

All the assumptions considered in the following are summarized in \tablename{}~\ref{table model 2}. Under these considerations, the link between $q$ and the couple of parameters $(\Delta r /r_{\rm ev}, \beta)$ is represented in \figurename{}~\ref{clues}. Because of its dependence on $2\pi \nu_{\rm max}/\mathcal{N}_{\rm b}$, the relation between the value of $q$ and $(\Delta r /r_{\rm ev}, \beta)$ is not unique in the transition regime; it is thus marginally indicated by a blue rectangle in \figurename{}~\ref{clues} for a large range of values for $\Delta r /r_{\rm ev}$ between $0.2$ and $0.6$. We emphasize that the uncertainties on the definition of the applicability domains of the Type-a, transition and Type-b regimes as a function of $q$ on the RGB will have no impact on the qualitative interpretation of the observed variations, since the behavior of $q$ with $\Delta r /r_{\rm ev}$ and $\beta$ follows the same global trend for the three regimes.

%------------------------------------
\subsubsection{Subgiant and young red giant stars}
%---------------------------------------

As an illustration, we consider the case of a subgiant star with typically $q=0.50$. For subgiant stars, the uncertainties on the measurements of $q$ by \cite{Mosser2017b} are lower than 20\%; we thus adopt a relative uncertainty of 10\% on its value.
As shown in \figurename{}~\ref{clues}, $q=0.50\pm 0.05$ is compatible with the ranges of parameters $\beta\in[1.2,1.5]$ and $\Delta r /r_{\rm ev} \in [0,0.25]$. At such a high value of $q$, the degeneracy with $\beta$ and $\Delta r /r_{\rm ev}$ leads to a large ambivalence on the determination of the structure. Furthermore, even if $\beta$ is fixed for instance at $1.3$, such value of $q$ can either correspond to $\Delta r/r_{\rm ev}=0.22\pm0.03$ or $\Delta r/r_{\rm ev}=0.03\pm0.02$, because the non-monotonic behavior of $\mathcal{G}_0$ with $\Delta r/r_{\rm ev}$. Without any other information than the value of $q$, it is therefore not possible to determine unequivocally the properties of the evanescent region in subgiant stars through a direct confrontation to the asymptotic analytical expression of $q$. 

Just after the end of the subgiant branch, we now consider as an example the case of a young red giant star with typically $q=0.20$.
For red giant stars, the automated measurement method of \cite{Mosser2017b} leads to large uncertainties on $q$. In contrast, the case-by-case study of \cite{Hekker2018} in few stars showed that the error bars on $q$ can reach at most $ 0.02$ on the RGB. We thus adopt an optimistic uncertainty of $0.02$ on the RGB.
Because of the ambiguity on the formalism to use in this case, both are compared. In the formalism of \cite{Takata2016a}, the model predicts $\Delta r /r_{\rm ev} = 0.48\pm 0.04$ for $q=0.20\pm 0.02$, with a small sensitivity regarding $\beta$. In contrast, in the formalism of \cite{Shibahashi1979}, the model predicts $\Delta r /r_{\rm ev}=0.28\pm0.07$ at $\beta=1.5$ and $\Delta r /r_{\rm ev}=0.35\pm0.08$ at $\beta=1.2$; in this case, the sensitivity of $q$ to $\beta$ cannot be neglected with the current uncertainties on $q$.
However, since the considered configuration lies somewhere between the limiting cases of a thick and of a very thin evanescent region, we thus cannot conclude on which prediction is the most relevant. We emphasize the same issue holds for the young subgiant stars with low observed values of $q$ between about $0.15$ and $0.25$.

Despite these difficulties related to the degeneracy of $q$ and the ambiguity on the coupling formalism to use, the model can bring us some clues on the structural origins of the global observed variations in $q$ during the evolution on the subgiant branch\footnote{In the youngest observed subgiant stars, with typically $\Delta \nu \sim 60~\mu$Hz and $q\sim 0.15$, the number of radial nodes inside the buoyancy cavity is close to about four \citep[e.g.,][]{Benomar2014}. Although the asymptotic approximation is questionable, we assume that the variations of $q$ at this stage can also be interpreted in the considered framework, at least qualitatively. This seems reasonable since the asymptotic expression of mixed modes fits the data in an acceptable way \citep{Mosser2017b}.}. In particular, the increase in $q$ from about $0.15$ to a peak value around $0.5$ at the beginning of the subgiant branch is necessarily associated with a predominant decrease in $\Delta r /r_{\rm ev}$, from about $0.50$ to at least $0.15$. The subsequent decrease in $q$ toward values lower than $0.2$ at the very beginning of the RGB automatically results in a predominant increase in $\Delta r /r_{\rm ev}$, toward about $0.5$. Moreover, the large spread in the values of $q$ around the short peak phase even suggests that $\Delta r /r_{\rm ev}$ reaches values lower than $0.15$. Indeed, the high sensitivity of $q$ to both parameters as well as its non-monotonic behavior with $\Delta r /r_{\rm ev}$ in this (very thin coupling region) regime may result, for a given stellar mass, in high amplitude variations over this short evolutionary phase, even if the structure slowly varies. Besides, for two stars with different stellar masses at a given value of $\nu_{\rm max}$, small differences in their mid-layer structures may also lead to significantly different values of $q$. The behavior of $\mathcal{G}_0$ in very thin evanescent regions thus represents a plausible explanation for the large spread in the values of $q$ measured by \cite{Mosser2017b} during the subgiant peak phase.

%--------------------------
\subsubsection{Transition phase and evolved red giant stars}
%-------------------------
\label{ER clues}

\begin{figure}
\centering
\includegraphics[width=\hsize,trim= 0.8cm 0cm 0.8cm 1cm, clip]{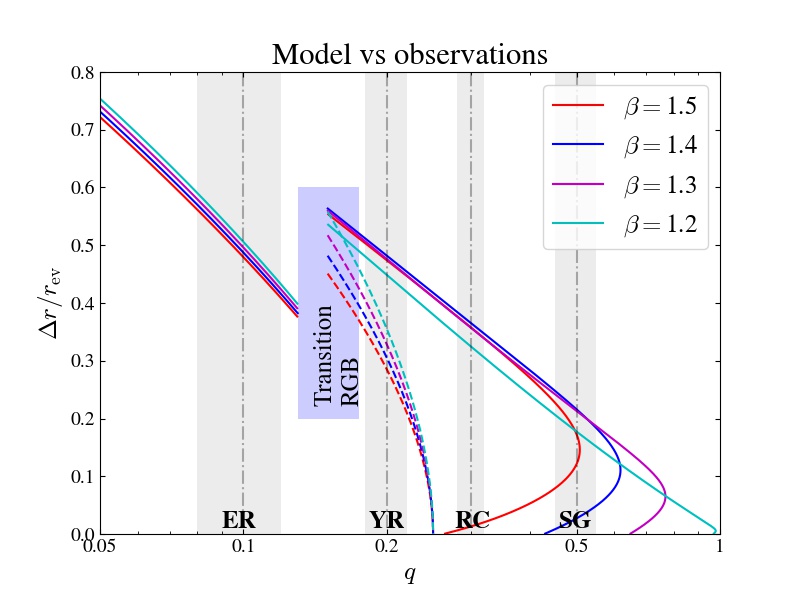} 
\caption{Normalized extent of the evanescent region as a function of $q$, for typical values of $\beta$. Solid lines represent the predictions of the model in the cases of Type-a very thin evanescent regions and thick Type-b evanescent regions for $q\gtrsim 0.15$ and $q\lesssim0.13$, respectively. The dashed lines correspond to the predictions for thick Type-a evanescent regions for $q\gtrsim 0.15$. The  blue rectangle indicates the estimated domain of the transition regime between Type-a and Type-b evanescent regions on the RGB. Typical values (vertical dash-dotted lines) and their error bars (vertical gray strips) are represented for subgiant (SG), young red giant (YR), evolved red giant (ER) and red clump (RC) stars (see main text).}
\label{clues}
\end{figure}

In the subsequent transition phase, we can consider as in \sectionname{}~\ref{transition} that variations in $\beta$ are small enough to not affect $q$. In the framework of the model, the decrease in $q$ can thus be associated with the predominance of either an increase in $\Delta r /r_{\rm ev}$, either the progressive migration of the evanescent region from the radiative to the convective region, or the simultaneous effects of these two ingredients.

In later stages, we can consider as an illustration the case of an evolved red giant stars with typically $q=0.10\pm0.02$. As seen in \figurename{}~\ref{clues}, such a value of $q$ is compatible with $\Delta r /r_{\rm ev}=0.48\pm0.08$ for $\beta=1.2$ or with $\Delta r /r_{\rm ev}=0.51\pm0.08$ for $\beta=1.5$. Given the current error bars on $q$, the sensitivity of $q$ to $\beta$ is thus small enough to be neglected at this evolutionary stage. Therefore, we confirm the prediction obtained in \sectionname{}~\ref{transition}; the slight decrease in $q$ observed by \cite{Mosser2017b} in the more evolved red giant stars of the sample results from an increase in $\Delta r /r_{\rm ev}$ as $\nu_{\rm max}$ decreases during evolution.

%-------------------
\subsubsection{Red clump stars}
%-------------------

In the red clump, where the helium fusion takes place in the core, we consider as an example a star with typically $q=0.3$ and an optimistic error bar close to $0.02$ \citep[e.g.,][]{Mosser2017b,Hekker2018}. According to \figurename{}~\ref{clues}, such a value of $q$ is compatible with $\Delta r/r_{\rm ev} \in [0,0.25]$ and $\beta \in[1.2,1.5]$. As in subgiant stars, the link with the internal properties remains ambivalent.
However, unlike subgiant stars, the small spread in the observed values suggests that the structure of the intermediate evanescent region little varies during the evolution in the red clump; otherwise, the high sensitivity of $q$ to the structure in the regime of very thin evanescent regions would lead to larger variations, as in subgiant stars.
This expectation seems reasonable since the red clump corresponds to a stable phase of the evolution where the helium-burning core produces enough energy to counterbalance the weight of the above layers, while the subgiant branch corresponds to a short unstable phase where the core collapses and the envelope dilates rapidly.

%----------------
\subsection{Convective boundary signature in evolved RGB stars}
%----------------

As shown in \sectionname{}~\ref{ER clues}, $q$ turns out to be mainly sensitive to the thickness of the evanescent region in evolved red giant stars, and hence to the radius of the base of the convective region. Indeed, the lower $r_{\rm b}$, the higher $\Delta r$ and the lower $r_{\rm ev}$, so that the higher $\Delta r /r_{\rm ev}$ and the lower $q$. The measurement of the coupling factor in such stars may therefore enable us to probe the boundary between the radiative and convective zones.

To discuss this point, we can consider as an example the case of a $1.2M_\odot$ evolved red giant star with typically $\nu_{\rm max}\approx 60~\mu$Hz, $\Delta \nu\approx 6~\mu$Hz, and $q=0.05$. Such a value of $q$ is compatible with the asymptotic value computed for the standard model ER in \figurename{}~\ref{propdiag}, or with the minimum value measured by \cite{Mosser2017b} at this evolutionary stage. To go further, we follow the picture used in \sectionname{}~\ref{transition}; we assume that $\alpha_{\rm b}=\mathcal{N}_{\rm b}/\mathcal{S}_{\rm b}$ remains equal to $0.5$ on the considered phase and fix the value of $\beta$ at $1.5$. As a result, $q$ only depends on the ratio $2\pi \nu_{\rm max}/\mathcal{N}_{\rm b}$, as represented in \figurename{}~\ref{q_Nb}. We remind that this ratio is expected to decrease during the evolution on the RGB.

Under these considerations, the analytical model requires $2\pi \nu_{\rm max}/\mathcal{N}_{\rm b} \approx 0.6$ to explain $q=0.05$; the knowledge of $q$ thus enables us to infer that $\mathcal{N}_{\rm b}/2\pi\approx 100~\mu$Hz for this star with $\nu_{\rm max}\approx 0~\mu$Hz. Using standard stellar models with the same input physics as described in \sectionname{}~\ref{dispersion}, \cite{Pincon2019} found that the transition to the Type-b regime occurs at $\nu_{\rm max,t} \approx 95~\mu$Hz for $1.2M_\odot$ stars. Since $2\pi\nu_{\rm max,t}\approx \mathcal{N}_{\rm b,t}$ at the transition, we have \smash{$\mathcal{N}_{\rm b}/\mathcal{N}_{\rm b,t}\approx 100/95\approx 1.05 $.} Therefore, in this example, the analytical model predicts that $\mathcal{N}_{\rm b}$ increases of about 5\% as $\nu_{\rm max}$ decreases from $\nu_{\rm max,t}\approx 95~\mu$Hz to about $60~\mu$Hz. Since \smash{$\mathcal{N}_{\rm b}^2\sim G M_r(r_{\rm b})/r_{\rm b}^3$} and hence is mainly sensitive to the variations in the factor $r_{\rm b}^{-3}$, we deduce that this slight increase in $\mathcal{N}_{\rm b}$ is directly related to a slight decrease in $r_{\rm b}$. This seems reasonable since, according to \cite{Pincon2019}, the transition to the Type-b regime happens just before the luminosity bump, that is, before the radius of the base of the convective region attains its minimum during the first dredge-up of the chemical elements. 

However, the inward migration of $r_{\rm b}$ is not the only possible reason for observing such a value of $q$ at $\nu_{\rm max}\approx 60 ~\mu$Hz. This can also be explained considering the value of $\mathcal{N}_{\rm b}$ fixed (i.e., $r_{\rm b}$ fixed). Indeed, for $\mathcal{N}_{\rm b}/2\pi$ to be remain about equal to $100~\mu$Hz, that is, in agreement with the value inferred according to the analytical model in the present example, we can choose a slightly higher value of $\nu_{\rm max,t}$ equal to about $100~\mu$Hz (i.e., such as $\mathcal{N}_{\rm b,t}\approx 100~\mu$Hz). As shown by \cite{Pincon2019}, this increase in $\nu_{\rm max,t}$, compared to the value estimated through standard stellar models such as these considered as examples in the previous sections (e.g, Figs.~\ref{propdiag}-\ref{HR}), can be physically produced by the inclusion of an adiabatic overshooting region below the base of the convective zone since this process decreases $r_{\rm b}$ and thus increases $\mathcal{N}_{\rm b}$ at a given value of $\nu_{\rm max}$. Moreover, the study of \cite{Khan2018} supports the need for overshooting in such evolved RGB models in order to match the location of the RGB luminosity bump with seismic observations. To estimate the quantity of overshooting needed to increase $\nu_{\rm max,t}$ (i.e,  $\mathcal{N}_{\rm b,t}/2\pi$) from $95~\mu$Hz to $100~\mu$Hz, we can proceed as follows. First, we emphasize that overshooting does not modify the formal shapes of $\mathcal{N}$ and $\mathcal{S}_1$, and thus the analytical model can still be applied. Second, at the base of the convective region of the considered evolved stars, it is reasonable to assume $J\approx 1$ (e.g., see \figurename{}~\ref{propdiag}). As a consequence, this high density contrast compared to helium core implies (see \appendixname{}~\ref{equivalence})
\algn{
\left(\deriv{\ln p}{\ln r}\right)_{r_{\rm b}} =  \frac{r_{\rm b}}{H_{\rm p,b}}\approx 6J-2\approx 4\; ,
\label{dpdr}
}
where $H_{\rm p,b}$ is the pressure scale height at $r_{\rm b}$. Then, assuming that the variation in $\mathcal{N}_{\rm b}$ mainly results from its dependence on $r_{\rm b}^{-3/2}$, using \eq{dpdr} and rewriting the amount of decrease in $r_{\rm b}$ caused by overshooting as $\delta r_{\rm b} = -\alpha_{\rm ov} H_{\rm p,b}$, with $\alpha_{\rm ov}$ the overshooting parameter, we obtain
\algn{
\frac{\delta \nu_{\rm max,t}}{\nu_{\rm max,t}} \approx \frac{\delta \mathcal{N}_{\rm b,t}}{\mathcal{N}_{\rm b,t}} \approx -\frac{3}{2}\frac{\delta r_{\rm b}}{r_{\rm b}}=\frac{3\alpha_{\rm ov}}{8} \;.
}
Therefore, with $\delta \nu_{\rm max,t}/\nu_{\rm max,t} \approx 0.05$, we find that $\alpha_{\rm ov}\approx 0.15$ is sufficient for the analytical expression to predict $q=0.05$ at $\nu_{\rm max}=60~\mu$Hz, without considering any variation in $\mathcal{N}_{\rm b}$ along with evolution. Interestingly, this is close to the overshooting value found by \cite{Khan2018} to fit the luminosity bump to observations in such stars.

In reality, the value of the coupling factor in such evolved red giant stars, that is, for $\Delta \nu \lesssim 10~\mu$Hz, is certainly the result of the simultaneous influences of the migration rate of $r_{\rm b}$ during the first dredge-up and the extent of overshooting. This simple estimate clearly emphasizes the potential of $q$ in such stars to provide constraints on the dynamical processes at the interface between the convective and the radiative regions, as for instance the convective mixing of the chemical elements.

%------------
\subsection{Frequency-dependence of $q$ over the spectrum}
%-----------
\label{frequency model}

When comparing the analytical model to observations in the previous sections, we implicitly assume that the current measurements of $q$ are representative of the predicted values at the frequency $\nu_{\rm max}$.
Indeed, the usual measurement methods assume $q$ as a constant parameter to fit, independent of the oscillation frequency around $\nu_{\rm max}$ \citep[e.g.,][]{Buysschaert2016,Mosser2017b,Hekker2018}. Nevertheless, previous works highlighted the frequency-dependence of $q$ in stellar models \citep[e.g.,][]{Jiang2014,Mosser2017b,Hekker2018}. From an observational point of view, $q$ can be considered frequency-independent if its variation over the observed frequency range of a given star is lower than the uncertainties on its value. In this section, we question the relevance of this assumption in the framework of the analytical model. In this model, we see in \sectionname{}~\ref{dependence} that $q$ depends on the $\alpha$ and $\beta$ parameters. While $\beta$ is supposed to be fixed over the evanescent region, $\alpha$, and thus $\Delta r /r_{\rm ev}$, can depend on the oscillation frequency since it depends by definition on the frequency-dependent turning points.

In subgiant, young red giant and red clump stars with Type-a evanescent regions, it is easy to show that $\alpha=\mathcal{N} (r)/\mathcal{S}_1 (r)$ according to \eq{power law N and S}{alpha def}. This is the consequence of the assumed parallel variations of $\mathcal{N}$ and $\mathcal{S}_1$. Unlike $r_1$ and $r_2$, $\alpha$ is thus frequency-independent, so that $q$ is frequency-independent too. The analytical model thus supports the assumption of the frequency-independence of $q$ in such stars \citep[e.g.,][]{Mosser2017b}.

In contrast, for RGB stars in the transition regime, we show in \sectionname{}~\ref{transition} that $q$ explicitly depends on the oscillation frequency (see \figurename{}~\ref{q_Nb}). In more evolved RGB stars with Type-b evanescent regions, we obtain $\alpha=[r_2(\sigma)/r_{\rm b}]^\beta$ according to \eq{alpha def}, since $r_1 (\sigma) \approx r_{\rm b}$. Using \eq{model B}, it turns out that $\alpha=\sigma/\mathcal{S}_{\rm b}$. The model therefore predicts that the coupling factor depends on the oscillation frequency in such RGB stars. The question is then to know whether its variation is significant over a typical observed frequency range around $\nu_{\rm max}$ or not. To answer this question, we can proceed as in \cite{Pincon2019} and assume that the observed oscillation spectrum of a given star belongs to a $6\Delta \nu$-wide frequency range centered around $\nu_{\rm max}$, as expected from theoretical estimates and observations \citep[e.g.,][]{Grosjean2014,Mosser2018}. Using the $\nu_{\rm max} - \Delta \nu$ scaling law in RGB stars \citep[e.g.,][]{Mosser2012b}, the observed frequency range is contained in $[\nu_{\rm max}-\delta \nu,\nu_{\rm max}+\delta \nu]$, where $\delta \nu \approx 3 \Delta \nu \approx 0.84 \nu_{\rm max}^{0.75}$. Considering a large range of values for $\nu_{\rm max}$ between $40$ and $250~\mu$Hz in such typical RGB stars (see \tablename{}~\ref{table model 2}), we approximately obtain $ 0.21~\nu_{\rm max} \lesssim \delta \nu \lesssim 0.33~ \nu_{\rm max}$. We thus consider a mean value of $\delta \nu\approx0.27~\nu_{\rm max}$ in the following. Therefore, using the theoretical framework presented in \sectionname{}~\ref{transition}, we find that $\delta q=\vert q(\nu_{\rm max}+\delta\nu)-q(\nu_{\rm max}-\delta\nu) \vert\gtrsim 0.02$ whatever the value of $\alpha_{\rm b}$ between $0.4$ and $0.6$, for $2\pi \nu_{\rm max}/\mathcal{N}_{\rm b} \in [0.4-2]$ (see \figurename{}~\ref{q_Nb}). This range of values for $2\pi \nu_{\rm max}/\mathcal{N}_{\rm b}$ corresponds to stars between the transition phase and one point just before the luminosity bump in the Type-b regime. It turns out that the difference in $q$ over the observed frequency range reaches a maximum value of about $\delta q\approx 0.05$, $0.07$ and $0.08$ at $2\pi \nu_{\rm max} /\mathcal{N}_{\rm b} \sim 1$, for $\alpha_{\rm b}=0.4,0.5$ and $0.6$, respectively. Given the current precision on the measurement of $q$ of about $0.02$ \citep[e.g.,][]{Hekker2018}, this result clearly demonstrates that the variation of $q$ with frequency in evolved red giant stars around the transition to the Type-b regime cannot be neglected. We emphasize that such a large variation over the observed frequency range is consistent with the variations in the asymptotic values of $q$ computed by \cite{Hekker2018} in RGB stellar models. It is also consistent with the frequency-dependence of $q$ that \cite{Cunha2019} had to consider to improve the fit of the asymptotic expression of mixed modes to the frequency pattern of RGB stellar models with typically $\nu_{\rm max}\lesssim 100 \mu$Hz, which correspond to stars with Type-b evanescent regions. 

In conclusion, while the developed model supports the frequency-independence hypothesis in subgiant, young red giant and red clump stars with Type-a evanescent regions, it demonstrates in contrast the need to account for the frequency-dependence of $q$ over the observed frequency range of red giant stars in the transition and in the Type-b regimes, that is, with about $5~\mu$Hz$\lesssim \Delta \nu \lesssim 18~\mu$Hz according to the estimate given in \tablename{}~\ref{table model 2}. In particular, this point may explain the too low asymptotic values of $q$ around $ 0.05$ computed in evolved stellar models compared to the values fit on theoretical and observed frequency spectra assuming $q$ as a frequency-independent parameter \citep{Mosser2017b,Hekker2018}. Further studies beyond this work are needed to understand how the frequency-dependence of $q$ can modify the previous measurements and their physical interpretation.

%--------------------------------
\section{Discussion}
%--------------------------------
\label{discussion}

%------------
\subsection{Possible improvements of the model}
%-----------
\label{limitations}

The analytical model that was developed in this work relies on several simplifying assumptions regarding the internal structure. In this section, we discuss the most critical points and the possible future improvements.

First, in Type-a evanescent regions, the model assumes that $\mathcal{N}$ and $\mathcal{S}_1$ vary as power laws of radius with the same exponent $\beta$. Under this hypothesis, the predicted value of $q$ is independent of the oscillation frequency in a given star. However, while \figurename{}~\ref{propdiag} shows that $\mathcal{N}$ and $\mathcal{S}_1$ vary as power laws of radius at first approximation, it also shows that they can have slightly different power-law exponents. This is actually related to the existing small gradient in the $J$ factor over the evanescent region (see \figurename{}~\ref{J graph}, \appendixname{}~\ref{equivalence} and the third paragraph of \appendixname{}~\ref{equivalence in conv} for details). As a result, the thickness of the evanescent region becomes dependent on the oscillation frequency; this feature certainly represents the reason for the frequency-dependence of $q$ observed in stellar models of subgiant, young red giant and red clump stars \citep[][]{Jiang2014,Mosser2017b,Hekker2018}. In this picture, the coupling factor will depend not only on the thickness of the evanescent region and the power-law exponent of $\mathcal{S}_1$, but also on that of $\mathcal{N}$. Besides, in Type-b evanescent regions, we demonstrate in \appendixname{}~\ref{equivalence in conv} that $\mathcal{S}_1$ cannot physically behave as a rigorous power law of radius, unlike what was assumed in the present work. In order to investigate analytically how the expression of $q$ is modified by such considerations in both types of evanescent regions, a first approach may consist in accounting for a first-order perturbation of the power-law structure and subsequently deducing the impact on the profiles of $\mathcal{S}_1$ and $\mathcal{N}$ in a consistent way by means of the continuity and hydrostatic equations (e.g., see \appendixname{}~\ref{constitutive} and \ref{equivalence in conv}).

Second, the model neglects the presence of a sharp \smash{$\mu$-gradient} observed at the base of the convective zone in evolved red giant stellar models. This sharp $\mu$-gradient is located in the deepest layer attained by the base of the convective region during the first dredge-up of the chemical elements on the RGB. Once the ascending hydrogen-burning shell encounters this layer, the resulting structural readjustment will produce the well-known RGB luminosity bump \citep[e.g.,][]{Kippenhahn2012}. The observational detection of the luminosity bump confirms the physical existence of such a feature in evolved red giant stars \citep[e.g.,][]{Khan2018}. This sharp $\mu$-gradient creates a spike in the Brunt-Väisälä frequency and a sharp jump in the Lamb frequency near the base of the convective region; this can be seen in the model ER of \figurename{}~\ref{propdiag}. These rapid variations can even be sharper in slightly less evolved models \citep[e.g., see Fig.~1 of][]{Mosser2017b}. It thus appears obvious that such features can have an impact on the coupling factor of mixed modes during the transition from a Type-a to a Type-b evanescent region. Indeed, close to this sharp $\mu$-gradient, the asymptotic approximation is not valid and the wave transmission factor through the evanescent region can be modified. Moreover, because of the resulting jump in the sound speed and the Lamb frequency, the outer turning point associated with the evanescent region can also be discontinuous as a function of the oscillation frequency, and so does the thickness of the evanescent region. We note that this feature represents, with the frequency-dependence of $q$, a plausible explanation for the discrepancies between the low asymptotic values computed in evolved stellar models and the observed values \citep{Mosser2017b,Hekker2018}. In order to investigate the effect of the sharp $\mu$-gradient on the coupling factor, future theoretical efforts will have to be made to go beyond the current asymptotic analyses, which are unable to deal properly with these sharp variations.

%------------
\subsection{Effect of stellar mass}
%------------
\label{stellar mass}

While the typical observed stellar masses range between $1M_\odot$ and $2M_\odot$, only $1.2M_\odot$ stellar models were considered as examples throughout this paper. In particular, the modeling of the evanescent region along with evolution was based on the internal structure of $1.2M_\odot$ stellar models. Since the properties of the stellar interior depends on stellar mass, the extension of the conclusions of the present work to the whole typical mass range needs to be discussed.

On the RGB, low-mass stars all follow the Hayashi line in the Hertzsprung-Russell diagram and thus share similar internal properties. It is easy to check through stellar models that the power-law behavior of $\mathcal{N}$ and $\mathcal{S}_1$, the high density contrast compared to the helium core and the transition from a Type-a to a Type-b evanescent region are also observed for typical masses between $1M_\odot$ and $2M_\odot$. The predictions and the interpretation of the variations in $q$ on the RGB thus hold valid  for the whole observed mass range.

On the subgiant branch, we checked using stellar models that stars with masses below about $1.4M_\odot$ again share similar properties to those depicted in the considered $1.2M_\odot$ models. Nevertheless, we observed in more massive subgiant stars that the deviation from the parallel variations of $\mathcal{N}$ and $\mathcal{S}_1$ discussed in \sectionname{}~\ref{limitations} is worsened; furthermore, the power-law assumption can even become questionable (at least, in the early subgiant phase). This appears to be related to a smaller density contrast over the evanescent region of such more massive stars. Indeed, we observed that the minimum value of the $J$ factor between the helium core and the base of the convective region is lower than, for instance, in the model SG (see \figurename{}~\ref{J graph}); as a result, the gradient of $J$ is higher and the departure from the power-law structure is more important (e.g., \appendixname{}~\ref{equivalence}). Interestingly, stars with masses higher than $1.4M_\odot$ correspond to stars whose mass of the helium core exceeds the Schönberg-Chandrasekhar limit (for a solar chemical composition) at the end of the main sequence. In such stars, the density contrast between the helium core and the envelope needs to rapidly increase on the subgiant branch to reach a new thermal equilibrium \citep[since the core has to contract while the envelope has to expand by mirror effect; e.g., see][]{Kippenhahn2012}. We thus suspect that the observed discrepancy compared to $1.2M_\odot$ stars is associated with the thermal instability of the helium core at the beginning of the subgiant branch in more massive stars. This may lead to a different relation between $q$ and the properties of the evanescent region in subgiant stars with masses higher than about $1.4M_\odot$ compared to that predicted in the present work. This issue deserves to be addressed in more detail in paper II by means of stellar models.

%------------
\subsection{Need for a medium coupling formalism}
%------------

\label{medium}

The available asymptotic analyses provide the expression for the coupling factor in two limiting cases only. On the one hand, \cite{Shibahashi1979} assumed that the turning points associated with the evanescent region are far away from each other. The wavenumber integral over the considered region is much higher than unity and the effect of the gradients in the equilibrium structure near the evanescent region can be neglected. This formalism can thus be applied to describe thick evanescent zones. On the other hand, \cite{Takata2016a} assumed that both turning points are very close to each other and accounted for the zeroth-order influence of the gradients in the equilibrium structure. More specifically, he assumed a particular ordering in Eq.~(50) of his paper, which reads $\Delta r /r_{\rm ev} \approx 2 \tanh ~(\tilde{a}~ K^{-1/2})$ in our notations, where $K$ and $\tilde{a}$ are a large real constant and a real function with a modulus lower than about unity, respectively. This formalism can thus be used to deal with very thin evanescent zones.

However, none of these formalisms formally describes the intermediate case of a medium coupling. As pointed out in \sectionname{}~\ref{expression type a}, the expression of \cite{Takata2016a}  does not even converge to the expression of \cite{Shibahashi1979} when the evanescent zone is artificially made very large. This dichotomy between both limiting cases and their unclear domains of applicability make ambiguous the comparison between asymptotic values of $q$ and observations, and demanded to use simplifying additional assumptions in \sectionname{}~\ref{interpretation}. New theoretical developments in the hypothesis of a medium coupling, that is, between both previous limiting cases, are therefore needed to overcome these  issues. Such improvements are currently investigated (Takata, priv. comm.).

%---------------
\section{Conclusions and prospects}
%---------------
\label{conclusions}

In the present work, we investigated the potential of the coupling factor of dipolar mixed modes, $q$, in probing the internal properties of post-main sequence stars. In this first paper of the series, an analytical approach, based on the asymptotic theory of mixed modes, was considered.

This seismic parameter is known to be sensitive to the structure of the evanescent coupling region of mixed modes. First, guided by stellar models, we highlighted that the evanescent region of typically-observed mixed modes progressively migrates from the radiative to the convective regions after the main sequence. In subgiant and young red giant stars, the evanescent region is located in the radiative layers between the hydrogen-burning shell and the base of the convective zone. There, the Brunt-Väisälä and Lamb frequencies vary at first approximation as power laws of radius with the same exponent. We clearly showed using the hydrostatic and the continuity equations that this peculiar behavior directly originates from the high local density contrast compared to the helium core. In evolved RGB stars, which correspond to stars with $\Delta \nu \lesssim 10~\mu$Hz, the evanescent region is situated in the lowest part of the convective region, where the Lamb frequency also varies at first approximation as a power law of radius but the Brunt-Väisälä frequency vanishes. Assuming such a power-law structure and accounting for the migration of the evanescent region with evolution, we obtained analytical expressions of the coupling factor based on the asymptotic formulations of dipolar mixed modes of \cite{Shibahashi1979} and \cite{Takata2016a}. The non-negligible effect of the perturbation of the gravitational potential on the coupling factor of dipolar modes was clarified and taken into account. 

The model shows that the value of $q$ depends in general on two structural parameters: first, the radial extent of the evanescent region normalized by its middle radius, denoted with $\Delta r /r_{\rm ev}$; second, the density scale height in the evanescent region, represented by the logarithmic slope of the critical frequencies, denoted with $\beta$. The model predictions provide us with some clues about the reasons for the variations of $q$ during evolution observed by \cite{Mosser2017b}:
\begin{itemize}
\item From the subgiant branch to the beginning of the RGB, that is, for $\Delta \nu \gtrsim 18~\mu$Hz, the peak in the observed value of $q$ is necessarily associated to a decrease in $\Delta r /r_{\rm ev}$ to at least about $0.15$, followed by a subsequent increase. The large spread in the measurements during the peak phase suggests that the coupling region can even become much thinner. \\

\item In the following transition phase, which we estimated for $10~\mu$Hz~$\lesssim \Delta \nu\lesssim 18~\mu$Hz, the evanescent region progressively migrates from the radiative to the convective regions. The small observed decrease in $q$ over this phase can result either from a predominant increase in $\Delta r /r_{\rm ev}$, either from the predominance of the progressive migration of the evanescent region, or from both ingredients simultaneously. Unlike the gravity offset, $\varepsilon_{\rm g}$, we find that the progressive migration of the evanescent region is not associated with a sharp variation in $q$. \\

\item In more evolved red giant stars, that is, for $\Delta \nu \lesssim 10~\mu$Hz, the coupling factor turns out to be mainly sensitive to $\Delta r /r_{\rm ev}$; the slight decrease in $q$ observed in such stars mainly results from an increase in $\Delta r /r_{\rm ev}$ as $\nu_{\rm max}$ decreases. \\

\item In red clump stars, the evanescent regions of which are in the radiative region, the negligible variations in $q$ around $0.3$ and the small spread in the measurements indicate that the structure of their mid-layers very slowly varies during the helium burning phase.
 \end{itemize}

The model also gives us more specific insights into the probing potential of $q$. In very thin evanescent regions, as found in subgiant and red clump stars, the degeneracy of $q$ prevents us from unambiguously concluding on the mid-layer properties without any other information than its value.
In contrast, in evolved RGB stars, $q$ can provide us with stringent constraints on the location and migration of the base of the convective region, as well as on the local convective extra-mixing. This appears to be promising for instance for the study of the first dredge-up or the RGB luminosity bump. Nevertheless, we find that a relevant comparison between the theoretical predictions and observations in these evolved stars needs to account for the frequency-dependence of the coupling factor over the observed frequency range, which is not considered in the usual measurement methods and could affect the physical interpretation of the observations. Finally, the model also highlights the need for the development of a medium coupling formalism, tackling the intermediate case between very thin and thick evanescent regions. This will enable us to assess properly and unequivocally the relation between $q$ and the internal properties, in particular in young subgiant and young red giant stars; this point will demand to go beyond the current theoretical picture of mixed modes.

All the lessons already learned from the analytical model will help us go further in the investigation. The next step will consist in exploring the link between the asymptotic value of $q$ and the internal properties using stellar models of different masses, and confronting the results to the observations. This will permit to question the results of the present study, as well as the structural origin of the variations in $q$ in subgiant and red clump stars, for which the degeneracy of $q$ with the internal properties prevents us from directly concluding. This will be subject of the forthcoming paper II of this series.
However, in order to test stellar models, it is also necessary to develop seismic diagnoses that can independently probe the evanescent region, as for instance the diagnosis developed by \cite{Goupil2013} for the internal rotation. Another seismic index is thus necessary for making such a model-independent probing of the mid-layer structure possible. The simultaneous measurement of the gravity offset of mixed modes with $q$ represents an interesting option. Actually, \cite{Takata2016a} demonstrated that $\varepsilon_{\rm g}$ also depends on $\Delta r /r_{\rm ev}$ and $\beta$. Moreover, in evolved RGB stars, \cite{Pincon2019} showed that $\varepsilon_{\rm g}$ is also sensitive to the base of the convective region and overshooting. Therefore, both measurements of $q$ and $\varepsilon_{\rm g}$ could be complementary and provide a model-independent information on the mid-layer structure of red giants. This promising possibility will be studied in a further paper.

%%%%%%%%%%%%%%%%%%%%
\begin{acknowledgements}
C. P. acknowledges Masao Takata for the fruitful and helpful discussions about mixed modes, for his careful reading of the paper, as well as for his support. We also warmly thank Benoît Mosser for his valuable remarks on the paper. During this work, C. P. was partially funded by a postdoctoral fellowship from F.R.S.-FNRS (Belgium) and from CNRS (France). 
\end{acknowledgements}

%%%%%%%%%%%%%%%%%%%%%
\bibliographystyle{aa} 
\bibliography{bib}

%%%%%%%%%%%%%%%%%%%%%
\appendix

%__________________
%
\section{Impact of the Cowling approximation on dipolar mixed modes}
%__________________
\label{Cowling}

The Cowling approximation consists in neglecting in the stellar oscillation equations all the terms associated with the Eulerian perturbation of the gravitational potential, denoted by $\Phi^\prime$ \citep[e.g.,][]{Cowling1941,Unno1989}. This hypothesis is actually equivalent to assume $\rho \nabla_r \Phi^\prime\ll \rho^\prime g$ in the radial linearized momentum equation, where $\nabla_r$ is the radial component of the nabla operator, $\rho$ is the density, $g$ is the gravitational acceleration and $\rho^\prime$ represents the Eulerian perturbation of density. Indeed, it is possible to show that this latter assumption implies $\rho \Phi^\prime \ll p^\prime$ in the horizontal linearized momentum equation, where $p^\prime$ represents the Eulerian perturbation of pressure. As a result,  the linearized Poisson equation, $\nabla^2 \Phi^\prime = 4\pi G \rho^\prime$, becomes decoupled from the momentum equation, which greatly simplifies the stellar oscillation equations. In this section, we discuss the validity of the Cowling approximation and in particular, its effect on the physical description of dipolar mixed modes in evolved stars.

%-----------------------------
\subsection{Validity conditions}
%-----------------------------
\label{Cowling A}

In the following, we assume $\nabla_r \Phi^\prime \sim \Phi^\prime / \lambda_r$ in order of magnitude. In the regions where the WKB approximation is met, $\lambda_r$ represents either the local radial wavelength or the length of decay of the wave function, depending on whether we consider oscillating or evanescent regions, respectively. Otherwise, it is on the order of the pressure scale height, $H_{\rm p}$.
In these considerations, the Poisson equation can be rewritten given the decomposition onto the spherical harmonics such as
\algn{
\left(\frac{1}{\lambda_r^2}+k_h^2\right)\Phi^\prime \sim 4 \pi G \rho^\prime  \; ,
\label{poisson}
}
where $k_h^2=\ell(\ell+1)/r^2$ is the squared horizontal wavenumber with $\ell$ the angular degree. To go further, we can also rewrite the gravitational acceleration such as \smash{$g = G M_r/r^2= (4 \pi G r/ 3)~ \rho_{\rm av}$}, where $\rho_{\rm av}$ is the mean density inside the shell of radius $r$, which satisfies the condition $\rho_{\rm av}(r) > \rho(r)$. Using the latter expression of $g$, $\nabla_r \Phi^\prime\sim \Phi^\prime/\lambda_r$ and \eq{poisson}, orders of magnitude give
\algn{
\left(\frac{\rho\nabla_r \Phi^\prime}{\rho^\prime g} \right)\sim 3 \frac{\rho}{\rho_{\rm av}}   \frac{(\lambda_r/r)}{1+\ell(\ell+1)~ (\lambda_r /r)^2} \; .
\label{cowling}
}
Equation~(\ref{cowling}) tells us that the Cowling approximation is appropriate when at least one of three following conditions is met:
\begin{itemize}
\item[(A)] the considered angular degree $\ell$ is higher than unity;
\item[(B)] the short-wavelength WKB approximation is met, that is, if \smash{$\lambda_r/r \ll1$;}
\item[(C)] or the density contrast between the considered layer and the inner bulk is high, that is, if $\rho /\rho_{\rm av} \ll 1$ (i.e., $J\sim 1$).
\end{itemize}
The conditions (A) and (B) are related to an averaging effect over the volume of the star. Indeed, if the number of nodes in the horizontal or in the radial directions is large enough, the small-scale density perturbations integrated over the whole volume of the star are expected to have only a small effect on the local gravitational acceleration. The condition (C) is related to the central character of the gravitational interaction. If \smash{$\rho /\rho_{\rm av} \ll 1$}, which is equivalent to the limit of $J \rightarrow 1$ according to \eq{def J}, the layers below the shell of radius $r$ can be regarded as a massive central point. The local gravitational acceleration thus remains mostly radial and quasi-unaffected compared to the equilibrium state. Actually, this confirms through orders of magnitude the result of \cite{Takata2016a} who rigorously demonstrated for dipolar mixed modes that the oscillation equations accounting for the effect of $\Phi^\prime$ tends to the equations within the Cowling approximation in the limit of $J\rightarrow 1$, as mentioned in \sectionname{}~\ref{dispersion}.

In the case of dipolar modes, the first condition is not met since the horizontal nodes are concentrated on only one equatorial plane. 
Nevertheless, the second and the third conditions can still be met depending on the regions.

%-----------------------------
\subsection{Effect on dipolar modes in evolved stars}
%-----------------------------
\label{Cowling B}

Well inside both resonant cavities of evolved stars, that is, far enough away from the associated turning points, the WKB approximation is met and the Cowling hypothesis is consequently valid. This point can be more thoroughly discussed in the light of the analysis of dipolar mixed modes by \cite{Takata2016a}. In the inner buoyancy cavity where $\sigma^2\ll (\mathcal{N}^2$ and $\mathcal{S}_1^2$), the dispersion relation for dipolar modes in \eq{k_r} becomes
\algn{
k_r^2 =  \frac{2N^2}{ r^2 \sigma^2}  \left[ 1+O\left(\frac{\sigma^2}{\mathcal{N}^2}\right)\right]\approx\frac{2N^2}{ r^2 \sigma^2}  \; ,
\label{k_r g}
}
while in the external acoustic cavity where \smash{$\sigma^2\gg (\mathcal{N}^2$ and $\mathcal{S}_1^2$),} it reads
\algn{
k_r^2= \frac{\sigma^2 }{ c^2} \left[ 1+O\left(\frac{\mathcal{S}_1^2}{\sigma^2}\right)\right] \approx  \frac{\sigma^2 }{ c^2}\; .
\label{k_r p}
}
As shown by \eq{k_r g}{k_r p}, the dispersion relation is not modified at leading order by the Cowling approximation in such regions. The correction resulting from the effect of $\Phi^\prime$ (i.e., through the $J$ factor) appears only in higher order terms. This explains why the asymptotic expressions of the period spacing and the frequency large separation of dipolar mixed modes, which correspond to leading-order terms in the asymptotic formulation, are not modified within the Cowling approximation, unlike the gravity offset that corresponds to higher order terms \citep{Takata2016a,Pincon2019}.
We also note that \eq{k_r g} confirms in the case of dipolar oscillations the result of \cite{Dintrans2001}, who demonstrated that the Cowling approximation is valid for low-frequency gravity waves of any angular degree because of their incompressible character (i.e., when $\sigma^2\ll N^2$). 

In contrast, inside the evanescent region and in the vicinity of the associated turning points, the WKB approximation does not apply since $\lambda_r \sim H_{\rm p}$. First, this is easy to confirm close to the turning points since $k_r^2 \rightarrow 0$. Second, inside the evanescent zone, \eq{k_r} shows that the squared modulus of the radial wavenumber is equal to 
\algn{
\vert k_r^2\vert =  \frac{2 J^2}{r^2} \left(1-\frac{\mathcal{N}^2}{\sigma^2} \right) \left(1-\frac{\sigma^2}{\mathcal{S}_1^2} \right) \lesssim \frac{2 J^2}{r^2}  \; .
\label{k_r ev}
}
As a result, the Cowling approximation is valid if the third condition about the density contrast (i.e., the limit of $J\rightarrow 1$) is met in this intermediate region. Figure~\ref{propdiag} shows that the value of $J$  can be low enough for the thickness of the intermediate coupling region to be significantly changed compared to the case within the Cowling approximation. Moreover, according to \eq{k_r ev}, the value of the wavenumber inside the evanescent region is also modified. Therefore, the value of the coupling factor must also be affected. We note that the younger the star, the higher the correction induced by $\Phi^\prime$, as the density contrast between the core and the envelope (and thus the $J$ factor) is lower, as shown in Figs.~(\ref{propdiag}) and (\ref{J graph}).

To summarize, our analysis shows that the Cowling approximation can lead to incorrect description of the coupling of mixed modes. The formulation of \cite{Takata2006b} in the non-Cowling case is mandatory for the study of the coupling factor of dipolar mixed modes, in particular for subgiant and young red giant stars.

%__________________________________

\section{On the structure of the evanescent regions in evolved stars}
%__________________________________
\label{on evanescent structure}

As seen in \figurename{}~\ref{propdiag}, the (modified) Brunt-Väisälä and Lamb frequencies behave at first approximation as power laws of radius in the evanescent region of evolved stars. In this section, we demonstrate that the origin of such a behavior is directly due to the high density contrast existing between the helium core and the envelope, which is measured by the $J$ quantity defined in \eq{def J}.

%---------------------------------------
\subsection{Link between $J$ and the critical frequencies}
%---------------------------------------
\label{equivalence relation}

In a first step, we show that the logarithmic derivatives of $N$ and $S_\ell$, denoted with $\beta_{\rm N}$ and $\beta_{\rm S}$, respectively, as well as the $J$ factor are linked to each other through the hydrostatic equilibrium and the continuity equations. We then demonstrate that an equivalence relation between these quantities exists when $\beta_{\rm N}$ and $\beta_{\rm S}$ are equal constants. The result is extended to the modified definitions $\mathcal{N}$ and $\mathcal{S}_1$ in the case of dipolar modes. The following statements only assume that $N$ and $S_\ell$ are decreasing functions of radius and that the first adiabatic index $\Gamma_1$ is constant. The case of radiative regions, in which $N^2>0$, is considered before discussing the case of convective regions, in which we assume $N^2 = 0$.

%-------------------
\subsubsection{Constitutive equations in radiative layers}
%------------------
\label{constitutive}
In this section, we aim to link $J$, $\beta_{\rm N}$ and $\beta_{\rm S}$ through the hydrostatic equilibrium and the continuity equations.
First, according to \eq{def J}, which results from the continuity equation, we obtain
\algn{
\deriv{\ln M_r}{x} = 3(1-J) \; ,
\label{M_r=f(J)}
}
where $x=\ln r$. 
Then, by differentiating $J$ with respect to $x$ and using again \eq{def J}, the derivative of $\rho$ with respect to $x$, which is renamed as $-\mathcal{B}(x)$, can be expressed as a function of $J$ only, that is,
\algn{
\mathcal{B}(x)\equiv-\deriv{\ln \rho}{x}=3J+\frac{1}{1-J}\deriv{J}{x} \; .
\label{B=f(J)}
}
To go further, we consider the hydrostatic equilibrium equation, that is, $(\dd p / \dd x) = - \rho g r$, where $g=G M_r/r^2$ is the gravitational acceleration. Dividing this equation by the product $-\Gamma_1 p$, taking the logarithmic derivative, and using \eq{M_r=f(J)}{B=f(J)}, we obtain
\algn{
\deriv{\ln\mathcal{V}}{x}&=\deriv{\ln \rho}{x}+\deriv{\ln M_r}{x}-1+\Gamma_1 \mathcal{V} \nonumber\\
&=-\mathcal{B}+3(1-J)-1+\Gamma_1 \mathcal{V}\; ,
\label{hydrostatic vs J}
}
where $\Gamma_1$ was considered as constant and we defined the positive quantity
\algn{
\mathcal{V}=-\frac{1}{\Gamma_1} \deriv{\ln p}{x}\; .
\label{def V}
}

The final step consists in rewriting the logarithmic derivatives of the Lamb and Brunt-Väisälä frequencies as a function of $J$, $\mathcal{V}$ and $\mathcal{B}$ only. First, the squared Lamb frequency (for any angular degree $\ell$) can be rewritten from the hydrostatic equilibrium as 
\algn{
S_\ell^2=\frac{\ell(\ell+1) c^2}{r^2}=\frac{\ell(\ell+1) \Gamma_1 p}{\rho r^2} \frac{ \rho g}{-(\dd p/\dd r)}
=\frac{\ell(\ell+1)  }{  \mathcal{V}}\frac{g}{r} \; . \label{S_l g/r}
}
Using the fact $\dd \ln( g/r)/\dd x=-3J$ according to \eq{M_r=f(J)}, the logarithmic derivative of the Lamb frequency is given by
\algn{
\beta_{\rm S}(x)\equiv - \deriv{\ln S_\ell}{x} =\frac{1}{2} \left(3J+\deriv{\ln \mathcal{V}}{x} \right) \; .
\label{eq alpha}
}
In a similar way, the logarithmic derivative of the Brunt-Väisälä frequency is equal, according to \eq{BVaisala} with $N=J\mathcal{N}$, to
\algn{
\beta_{\rm N}(x)\equiv - \deriv{\ln N}{x} = \frac{1}{2} \left( 3J - \deriv{\ln \mathcal{A}}{x}\right) \; , \label{eq beta_N}
}
where
\algn{
\mathcal{A}=\left(\frac{1}{\Gamma_1}\deriv{\ln p}{x}-\deriv{\ln \rho}{x} \right) =-\mathcal{V}+\mathcal{B}\; .
\label{def A}
}
We emphasize that $\mathcal{A}$ is a positive quantity in the radiative zone since $N^2=g\mathcal{A}/r \gtrsim 0$.

To derive a differential system linking $J$, $\beta_{\rm S}$ and $\beta_{\rm N}$, we first express $\mathcal{V}$ from \eq{eq alpha} such as
\algn{
\mathcal{V}(x) = \mathcal{V}_0 \exp\left( \int_{x_0}^x [2\beta_{\rm S}-3J]~\dd x \right) \; ,
\label{V vs beta_S}
}
where the subscript 0 denotes the value at $r=r_0$ (i.e., at $x=x_0$). Then, by integrating \eq{eq beta_N}, we also obtain
\algn{
\mathcal{A}(x)=-\mathcal{V}+\mathcal{B}=\mathcal{A}_0
\exp \left( -\int_{x_0}^x [2\beta_{\rm N}-3 J ]~ \dd x \right) \; .
\label{A vs beta_N}
}
By injecting \eq{V vs beta_S} into \eq{hydrostatic vs J}{A vs beta_N}, the $\mathcal{B}$ quantity in \eq{B=f(J)} can be expressed following three different ways as a function of the dependent variables $J$, $\beta_{\rm S}$ and $\beta_{\rm N}$, which are
\algn{
\mathcal{B}(x)&= 3J+\frac{1}{1-J}\deriv{J}{x} \label{B 1}\\
&=-2\beta_{\rm S} + 2 +  \Gamma_1 \mathcal{V}_0 \exp\left( \int_{x_0}^x [2\beta_{\rm S}-3J]~ \dd x \right) \label{B 2}\\
&=\mathcal{A}_0
\exp \left( -\int_{x_0}^x [2\beta_{\rm N}-3 J ]~ \dd x \right)\nonumber\\
&~~+\mathcal{V}_0 \exp\left( \int_{x_0}^x [2\beta_{\rm S}-3J]~ \dd x \right) \; .
\label{B 3}
}
Defining the dependent variables $W,$ $X,$ $Y$ and $Z$ such as
\algn{
W&=\mathcal{A}_0
\exp \left( -2\int_{x_0}^x \beta_{\rm N}~ \dd x \right)~~~~\; , ~~~~X= \mathcal{V}_0 \exp\left(2 \int_{x_0}^x \beta_{\rm S} ~\dd x\right)\; , \nonumber \\
Y&=\exp\left( \int_{x_0}^x 3J~ \dd x \right)~~~~\; , ~~~~Z=\deriv{Y}{x}=3J Y \; ,
\label{variables}
}
\eqss{B 1}{B 3}~can be rewritten in the form of a first-order ordinary nonlinear system of three differential equations, that is,
\algn{
\deriv{}{x}
\begin{pmatrix}
 X \\ Y \\ Z
\end{pmatrix}
=
\begin{pmatrix}
~~ 2 X+(\Gamma_1-1)~ \dfrac{X^2 }{Y}-WXY ~~\\ Z \\ ~~2Z^2Y+3Z-3W+WZY-\dfrac{X}{Y}~\left(\dfrac{3}{Y}-Z\right) ~~
\end{pmatrix} \; ,
\label{system}
}
with $W_0=\mathcal{A}_0$, $X_0= \mathcal{V}_0$, $Y_0=1$ and $Z_0=3J_0$ as initial values. Therefore, given the definitions of $W$, $X$, $Y$ and $Z$ in \eq{variables}, if at least one of the three variables $J$, $\beta_{\rm S}$ and $\beta_{\rm N}$ is specified around $r_0$, it is possible to deduce the two other ones through \eq{system}. Since the coupling factor of mixed modes mostly depend on $N$, $S_\ell$ and $J$ (see \sectionname{}~\ref{realistic structure}), its expression thus must account for this relation in order to be consistent. 

%-------------------------
\subsubsection{Expression of $J$ with given $\beta_{\rm S}$ and $\beta_{\rm N}$}
%-------------------------
\label{exp J}

Assuming that both $\beta_{\rm S}$ and $\beta_{\rm N}$ are known functions of $r$, the $J$ factor can actually be analytically expressed in a direct way. Indeed, by multiplying \eq{B 2}{B 3}~by the positive variable $Y$, we obtain the following quadratic equation for $Y$,
\algn{
A Y^2+BY+C=0 \; ,
}
with
\algn{
A&= \mathcal{A}_0 \exp \left( -2\int_{x_0}^x \beta_{\rm N} ~\dd x \right) > 0\\
B&=2(\beta_{\rm S}-1) \\
C&= -\mathcal{V}_0 \left[\Gamma_1-1 \right] \exp\left( 2 \int_{x_0}^x \beta_{\rm S}~ \dd x \right) < 0 \; .
}
The sign of $C$ results from the thermodynamical condition \smash{$\Gamma_1>1$.} The discriminant of the second-order equation, $\Delta = B^2-4AC$, is strictly positive and greater than $B^2$. The unique (positive) solution is thus given by \smash{$Y=(-B+\sqrt{\Delta})/2A$}. Computing the logarithmic derivative of $Y$, we then obtain
\algn{
J=\frac{2}{3}\left[ \beta_{\rm N}-\frac{1}{AY}\left( \frac{\beta_{\rm S}^\prime}{2}-\frac{\beta_{\rm S}^\prime(\beta_{\rm S}-1) + AC~[\beta_{\rm N} - \beta_{\rm S}]}{\sqrt{\Delta}}\right)\right] \; ,
\label{J=f(beta)}
}
where $\beta_{\rm S}^\prime(x) = (\dd \beta_{\rm S} / \dd x)$.

%-------------------------
\subsubsection{Equivalence relation}
%-------------------------
\label{equivalence}
\paragraph{Analysis:}

If $\beta_{\rm S}$ and $\beta_{\rm N}$ are both fixed to a given constant $\beta$, then \eq{J=f(beta)} shows that $J$ is uniform with respect to radius and equal to $2 \beta/3$. According to \eq{B=f(J)}, this assumption thus implies that the density varies as a power law with respect to radius, that is, such as $\mathcal{B}=-(\dd \ln \rho/\dd x)=3J$. Moreover, \eq{eq alpha} firstly shows that $(\dd \ln \mathcal{V}/\dd x)$ is constant; secondly, \eq{hydrostatic vs J} shows that this constant is zero and that $-(\dd \ln p/\dd x)=6J-2$. It is easy to check that this solution also satisfies \eq{B 1}. Therefore, the pressure also varies as a power law with respect to radius. In other words, the equation of state has a polytropic form such as $P \propto \rho ^\gamma$, where the polytropic exponent is equal to $\gamma=2-2/(3J)$. In conclusions, a necessary condition for $\beta_{\rm S}$ and $\beta_{\rm N}$ to be equal constants is that $J$ is constant and the equation of state is polytropic.

\paragraph{Synthesis:}

If $J$ is uniform and $P \propto \rho^\gamma$, $(\dd \ln \rho /\dd x)$ and $(\dd \ln p /\dd x)$ are also uniform according to \eq{B=f(J)}. Therefore, we deduce according to \eq{def V}{def A}~that $\beta_{\rm S}$ and $\beta_{\rm N}$ are both constants equal to $3 J / 2$, and that \smash{$\gamma=2-2/(3J)$.}

\paragraph{Conclusions:} The Brunt-Väisälä and Lamb frequency follow decreasing power laws of radius with the same exponent $\beta$ if and only if $J$ is uniform and the equation of state is polytropic. The equivalence configuration corresponds to a power-law solution of the internal structure, where $(\dd \ln p /\dd x)$, $(\dd \ln \rho /\dd x)$ and $(\dd \ln M_r /\dd x)$ are constants. As a corollary, $J$, $\beta$ and $\gamma$ are related to each other through the equations $\beta=3J/2$, as previously found by \cite{Takata2016a}, and \smash{$\gamma=2-2/(3J)$.}

%---------------------
\subsubsection{Extension to $\mathcal{N}$ and $\mathcal{S}_1$ for dipolar modes}
%---------------------
\label{equivalence b}

We now consider the modified Brunt-Väisälä and Lamb frequencies for dipolar modes, the logarithmic derivatives of which are denoted with $\beta_{\rm N}^\star$ and $\beta_{\rm S}^\star$. 

On the one hand, if $J$ is supposed to be uniform and $P \propto \rho ^\gamma$, it is then straightforward to show that $\beta_{\rm N}^\star=\beta_{\rm S}^\star=3J/2$ following the same reasoning as in the second paragraph of \sectionname{}~\ref{equivalence} and the definitions of $\mathcal{N}$ and $\mathcal{S}_1$ in \eq{NSJ}.

On the other hand, using $\beta_{\rm N}=\beta_{\rm N}^\star+(\dd \ln J/\dd x)$ and \smash{$\beta_{\rm S}=\beta_{\rm S}^\star-(\dd \ln J/\dd x)$} in \eq{B 1}{B 3},~and defining the new variables $W^\star$ and $X^\star$ by replacing $\beta_{\rm N}$ in \eq{variables} with $\beta_{\rm N}^\star$ and $\beta_{\rm S}$ with $\beta_{\rm S}^\star$, it is possible to demonstrate that  $W^\star$, $X^\star$, $Y$ and $Z$ are also linked by a first-order nonlinear system of three differential equations. If we impose that $\beta_{\rm N}^\star$ and $\beta_{\rm S}^\star$ are equal to a same constant, the differential system reduces to a nonlinear first-order system of two differential equations for $Y$ and $Z$. The constant $\beta=\beta_{\rm N}^\star=\beta_{\rm S}^\star$ is linked to the derivatives of $W^\star$ and $X^\star$, so that its choice is not arbitrary but is determined by the initial values of the problem. Actually, using the relation $\mathcal{B}=\mathcal{A}+\mathcal{V}$, it is easy to show that a trivial solution of the differential system given by \eq{B=f(J)}{hydrostatic vs J}{eq alpha}{eq beta_N}, in which we use $\beta_{\rm S}=\beta_{\rm S}^\star-(\dd \ln J/\dd x)$ and $\beta_{\rm N}=\beta_{\rm N}^\star+(\dd \ln J/\dd x)$, is also $J=J_0$. This condition again implies $\beta =3J_0/2$. Assuming that the Cauchy-Lipschitz theorem on the initial value problem is applicable over the evanescent region, which is usually the case, the constant solution $J=J_0$ is the unique solution. Moreover, since we also have $\beta_{\rm N}=\beta_{\rm N}^\star$ and $\beta_{\rm S}=\beta_{\rm S}^\star$ in this case, the equation of state also turns out to be polytropic according to the first paragraph of \appendixname{}~\ref{equivalence}.

Therefore, the same conclusions are reached for the modified critical frequencies; $\beta_{\rm N}^\star$ and $\beta_{\rm S}^\star$ are equal constants if and only if $J$ is uniform and the equation of state is polytropic, which implies \smash{$\beta=3J/2$}, \smash{$\gamma=2-2/(3J)$}, and a power-law configuration for the internal structure.

%---------------------
\subsubsection{Case of convective regions}
%---------------------
\label{equivalence in conv}

In the case of an adiabatic convective zone, we have $\mathcal{A}\approx 0$ since $P\propto \rho^{\Gamma_1}$. A general analytical expression of $J$ can thus be directly derived. Indeed, we have $\mathcal{V}\approx \mathcal{B}$ according to \eq{def A} with $\mathcal{A}\approx 0$. Using \eq{hydrostatic vs J}{eq alpha}, we can thus express $\mathcal{V}$ as a function of $\beta_{\rm S}$ and inject the result into \eq{hydrostatic vs J} to finally obtain
\algn{
J=\frac{2}{3} \beta_{\rm S} -\frac{\beta_{\rm S}^\prime}{3(\beta_{\rm S} -1)} \; .
\label{J conv}
}
Following the same reasoning as in \sectionname{}~\ref{equivalence}, it is possible to demonstrate that $J$ is uniform if and only if \smash{$\beta_{\rm S}$} is constant; in this case, \smash{$\beta_{\rm S}=3 J/2$}, the polytropic exponent is equal to \smash{$\gamma=2-2/(3J)$} and the internal structure behave as power laws of radius.

However, since the temperature gradient is adiabatic in convective zones, $\gamma=\Gamma_1= 5/3$ for a totally ionized monatomic ideal gas. If the equivalence was met, we would thus obtain $J=2$, which is not possible since $J<1$ by definition. Such an equivalence configuration is thus not valid in convective regions. Indeed, while the profile of the Lamb frequency behaves at first sight like a power law of radius in the evanescent region of the model ER in \figurename{}~\ref{propdiag}, we can see in \figurename{}~\ref{J graph} that the value of $J$ decreases of about 20\% over the evanescent region, and therefore cannot be regarded as rigorously uniform.

To understand the profile of $J$, it is useful to consider the derivative of $J$ with respect to radius. According to \eq{B=f(J)}, this reads
\algn{
\deriv{~J~}{\ln r}=(1-J)\left(-\deriv{\ln \rho}{\ln r}-3J \right) \; .
\label{deriv J}
}
As the value of $J$ rapidly varies from small values to about unity in the vicinity of the hydrogen-burning shell, the second term in brackets in the right-hand side of \eq{deriv J} becomes negative and $J$ starts decreasing with radius. In contrast, in the upper layers close to the surface, the density vanishes and $-(\dd \ln \rho / \dd r) \gg 1$. As a consequence, \eq{deriv J} becomes positive and $J$ starts increasing toward unity. These facts result in the existence of a minimum in the $J$ factor that is located in the external convective zone of the model ER (and in the model YR). We emphasize that, in the model SG, the minimum of $J$ is even located in the radiative region because of a lower density contrast\footnote{There is not such a local minimum in the solar case since the density contrast compared to the core is lower than in subgiant stars, so that $J$ is low except in the surface layers and monotonically increases with radius.}. Because of the necessary variation of $J$ in the convective region, the link between $\beta$ and the $J$ factor is thus provided by the differential equation in \eq{J conv}.

In the case of the modified expression of the Lamb frequency for dipolar modes, we note that the same conclusions hold true. The general expression linking $J$ and $\beta_{\rm S}^\star$ is however more complex than \eq{J conv} since $\beta_{\rm S}$ in this equation must be replaced by $\beta_{\rm S}^\star-(\dd \ln J/\dd x)$ and thus also involves the derivatives of $J$ with respect to radius.

%------------------------------
\subsection{Structure of the upper radiative layers}
%------------------------------
\label{upper rad}

In \sectionname{}~\ref{equivalence relation}, we have demonstrated that the Brunt-Väisälä and Lamb frequencies vary as power laws of radius with the same exponents between the hydrogen-burning shell and the base of the convective zone if and only if $J$ is uniform and the equation of state is polytropic. While it is easy to confirm through stellar models that these conditions are met at first approximation in this region of evolved stars \citep[e.g.,][see also Figs.~\ref{propdiag} and \ref{J graph}]{Maeder2009,Kippenhahn2012}, we propose in the following to discuss this point considering simplified analytical solutions of the internal structure equations. Actually, we show that these conditions mainly originate from the high density contrast between these layers and the helium core.

%-----------------------------
\subsubsection{Hypothesis of a high density contrast}
%-----------------------------
\label{polytropic approx}

In a first step, we assume that the density contrast between the considered layers and the helium core is very high. We then aim to show that such an assumption implies that the equation of state is polytropic and $J$ is uniform in a good approximation.
First, the definition of $J$ in \eq{def J} shows that the high density contrast compared to the helium core implies $\rho/\rho_{\rm av} \ll 1$ and $J\sim 1$. According to \eq{deriv J}, the derivative of $J$ with respect to radius is thus expected to be small so that $J$ remains about uniform, as expected.
Second, regarding the polytropic approximation, we need to use the following reasonable considerations to move forward:
\begin{enumerate}

\item The medium is non-degenerate and the equation of state can be assumed to follow the ideal gas law. In addition, the mean molecular weight can be supposed to be nearly uniform in this region if we neglect the impact of microscopic diffusion and the existence of local sharp variations resulting, for instance, from the migration of the base of the convective during evolution \citep[e.g.,][]{Cunha2015}.
\label{hyp 1} \\

\item The luminosity is constant and equal to the total luminosity of the star since it is fully produced in the underlying hydrogen-burning shell.\\

\item In the range of temperature expected in this region, the opacity predominantly can originate either from the Bremsstrahlung mechanism or even from bound-free electronic transitions near the base of the convective region \citep[e.g.,][]{Kippenhahn2012}. The Rosseland mean opacity, $\kappa_{{\rm op}}$, thus can be supposed to follow the Kramers law, that is, $\kappa_{{\rm op}}=\kappa_{{\rm op},0} \rho^r T^{-s}$, where $\kappa_{{\rm op},0}$ is a positive constant, $r\approx1$, and $s\approx 7/2$.
\label{hyp 3} \\

\item Owing to the high density contrast between the helium core and the envelope, the mass $M_r$ inside a spherical shell of radius $r$ is assumed to change very slowly with respect to radius, unlike the density $\rho$. It can thus be supposed to be equal to the mass of the helium core, $M_{\rm c}$, in the hydrostatic equilibrium equation.
\label{hyp 4} \\

\item Because of the ideal gas law, the temperature and the pressure are supposed to change rapidly with respect to radius in this region since the density is supposed to vary rapidly. %In more technical terms, their gradients must satisfy
%
%\algn{
%\left\vert\deriv{ \ln T}{\ln r} \right\vert\gtrsim~ \frac{1}{4+s+r} ~~~~\mathrm{and}~~~~ \left \vert \deriv{\ln p }{ \ln r} \right\vert \gtrsim~  \frac{1}{1+r} \; .
%}
%
\label{hyp 5} \\

\end{enumerate}
Then, using the hydrostatic equilibrium and the radiative transport equations, the pressure and the density are related to each other by
\citep[e.g., see Eq.~(21.16) of][]{Kippenhahn2012}
\algn{
p\approx p_{\rm i}\left( \frac{\rho}{\rho_{\rm i}}\right)^{(4+s+r)/(3+s)} \; ,
}
where the assumption~\ref{hyp 5} is considered in order to neglect the integration constants and the subscript i indicates values taken at radius $r_{\rm i}$, just above the hydrogen-burning shell.
Therefore,the equation of state is polytropic in a good approximation in the upper layers of the radiative region, with a polytropic exponent that is about equal to \smash{$\gamma=(\dd \ln p/\dd \ln \rho)=(4+s+r)/(3+s)=1.31$.} 

In consequence, assuming a high density contrast between the upper radiative layers and the underlying helium core is sufficient for the equation of state to be polytropic and 
the $J$ factor to be nearly uniform with radius in a good approximation. As shown in \appendixname{}~\ref{equivalence relation}, this is equivalent to a power-law behavior of the internal structure.

%-----------------------------
\subsubsection{Power-law solutions of the internal structure}
%----------------------------
\label{power-law}

Similarly to \cite{Applegate1988}, we now search for power-law solutions of the internal structure equations above the hydrogen-burning shell, that is, in the form of
\algn{
M_r&=M_{\rm i} \left( \frac{r}{r_{\rm i}}\right)^a ~~~~ , ~~~~ \rho=\rho_{\rm i} \left( \frac{r}{r_{\rm i}}\right)^{-b} \nonumber\\
p&=p_{\rm i} \left( \frac{r}{r_{\rm i}}\right)^{-c}  ~~~~,~~~~T=T_{\rm i} \left( \frac{r}{r_{\rm i}}\right)^{-d} \; ,
\label{power laws}
}
Using the assumptions \ref{hyp 1}-\ref{hyp 3} in \sectionname{}~\ref{polytropic approx} and injecting \eq{power laws} into the continuity, hydrostatic equilibrium, radiative transport equations and the ideal gas law, \cite{Applegate1988} found
\eqna{
a=&\dfrac{s-3r}{3+s-r}~~~~
&b=\dfrac{9+2s}{3+s-r}\\
c=&\dfrac{12+2s+2r}{3+s-r}~~~~
&d=\dfrac{3+2r}{3+s-r} \; .
}
Since the Bremsstrahlung effect and bound-free electronic transitions dominate the opacity, the power-law solutions are characterized by $a=1/11$, $b=32/11$, $c=42/11$, and $d=10/11$. As a result, the polytropic exponent is still equal to $\gamma=c/b=1.31$. This result justifies a posteriori the assumption \ref{hyp 5} made in \sectionname{}~\ref{polytropic approx} about the behavior of $P$ and $T$ with respect to radius. According to \eq{M_r=f(J)}, we also find \smash{$J=b/3=32/33$}, which is very close to unity. Therefore, this model suggests that the high density contrast between the considered layers and the helium core is not only sufficient, but also necessary to explain the power-law behavior of the internal structure.

%-----------------------------
\subsection{Concluding synthesis}
%----------------------------

In the region located between the hydrogen-burning shell and the base of the convective region, the high density contrast compared to the underlying helium core is the main responsible for the similar power-law behavior of the Brunt-Väisälä and Lamb frequencies. Indeed, under this condition, $J$ is uniform and the equation of state is polytropic, which implies that $p$, $\rho$ and $M_r$ are power laws with respect to $r$. As a result, the Brunt-Väisälä and Lamb frequencies are also power laws of $r$ with the same exponent, $\beta$. In addition, $\beta$ is linked to the $J$ factor by the simple relation $\beta \approx 3J/2$. In contrast, in the adiabatic convective envelope, such a configuration is forbidden: $J$ varies with respect to radius and is linked to the exponent $\beta$ through a more complicated differential equation.

%__________________________________
%
\section{Wave transmission and coupling factor through a parabolic barrier}
%__________________________________
\label{parabolic barrier model}

In this section, we study the problem of the wave transmission through a parabolic evanescent barrier. This case does not require any assumptions about the thickness of the evanescent region to be fully solved, contrary to the asymptotic analyses by \cite{Shibahashi1979} and \cite{Takata2016a} for realistic stellar structures (\sectionname{}~\ref{realistic structure}). We propose here to retrieve the expression of the wave transmission coefficient $T^2$ for this simplified case using the same method as \cite{Phinney1970} in a geophysical context.
In this situation, the wave equation to solve is provided by \eq{Psi eq}.

To continue, we use the change of variable
\algn{
\zeta^2= 4 \frac{\vert k_r(r_{\rm ev})\vert}{\Delta r}(r-r_{\rm ev})^2 \; , \label{cov zeta vs r}
}
so that \eq{Psi eq} can be rewritten in the form of the Weber's equation, that is,
\algn{
\derivs{\Psi}{\zeta} +\left(\frac{\zeta^2}{4}-a \right) \Psi = 0\; , \label{parabolic equation}
}
where the positive parameter $a$ is defined\footnote{In the case where $k_r^2$ is positive everywhere, there is no evanescent zone. The wave equation to solve is similar to \eq{parabolic equation}, except that $a<0$. In this case, incident waves still suffer partial reflection close to $r_{\rm ev}$; a coupling factor can therefore also be defined. This situation is equivalent to glitches in stars, as already discussed in \cite{Pincon2018}.} as
\algn{ a= \frac{1}{4} \vert k_r(r_{\rm ev}) \vert\Delta r \; .}
The general solution of \eq{parabolic equation} is given by a linear combination of the complex independent parabolic cylinder functions, namely $E(a,\zeta)$ and its complex conjugate $E^\star(a,\zeta)$ \citep[e.g.,][]{Abramowitz1972}. 
When $\vert \zeta^2 \vert \gg \vert a \vert$, that is, far enough from and on both sides of the region around $r =r_{\rm ev}$, the first-order asymptotic development of the parabolic cylinder function is given by
\algn{E(a,\zeta)\sim \sqrt{\frac{2}{\vert \zeta\vert}} \exp \left[+i\left( \frac{\zeta^2}{4}-a\ln\vert \zeta\vert+\frac{\pi}{4}+\frac{\Psi_2}{2}\right) \right] \: , \label{asymptotic E}}
where $\Psi_2=\arg \left[ \Gamma (1/2+ia)\right]$, $\Gamma$ is the gamma-function and $i$ is the imaginary unit. In the following, a time-dependence in $e^{i\sigma t}$ for the wave function and the positive branch in \eq{cov zeta vs r} are considered (i.e., such as $\zeta$ and $(r-r_{\rm ev})$ have the same sign everywhere). Within this convention, \eq{asymptotic E} shows us that $E(a,\zeta)$ corresponds to a regressive wave (i.e., traveling in the negative $r$-direction) where $\zeta >0$; it corresponds to a progressive wave (i.e., traveling in the positive $r$-direction) where $\zeta <0$. In contrast, $E^\star(a,\zeta)$ corresponds to a progressive wave where $\zeta >0$ and a regressive wave where $\zeta <0$.
For later purpose, it is helpful to note that the change of variable $\zeta^\prime =-\zeta$ does not modify the form of \eq{parabolic equation}, so that $E(a,-\zeta)$ and $E^\star(a,-\zeta)$ stand also for the regressive and  progressive independent solutions where $\zeta >0$, respectively, and inversely where $\zeta<0$.
Both sets of independent solutions $\{E(a,\zeta),E^\star(a,\zeta)\}$ and $\{E(a,-\zeta),E^\star(a,-\zeta)\}$ are linked by the following connection formula \citep[e.g.,][]{Abramowitz1972}
\algn{\sqrt{1+e^{2\pi a}}E(a,\zeta) - e^{\pi a}E^\star(a,\zeta)=iE^\star(a,-\zeta) \; . \label{connection formula}}

The transmission and reflection coefficients through the evanescent region can now be easily determined by means of this last relation. We consider an incident progressive wave coming from $\zeta=-\infty$ that is both reflected back and transmitted into the negative and positive $\zeta$-direction in the vicinity of the evanescent barrier. In the domain where $\zeta < 0$, the incident and reflected components are thus proportional to the progressive and regressive independent solutions $E(a,\zeta)$ and $E^\star(a,\zeta)$, respectively, that is, such as the wave function is equal to
\algn{
\Psi = I E (a,\zeta) + r E^\star(a,\zeta) \; ,
\label{phi inf}
}
where the constants $I$ and $r$ are the amplitudes of the incident and reflected waves, respectively.
In the domain where $\zeta>0$, the transmitted wave is proportional to the progressive solution $E^\star(a,-\zeta)$, that is, such as the wavefunction can be written
\algn{
\Psi=t E^\star(a,-\zeta)=-i t \left( \sqrt{1+e^{2\pi a}}E(a,\zeta) - e^{\pi a}E^\star(a,\zeta) \right)\; ,
\label{phi sup}
}
where the constant $t$ is the amplitude of the transmitted wave and where \eq{connection formula} was used in the second equality. Since the expressions of the wavefunction in \eq{phi inf}{phi sup}~are valid whatever the value of $\zeta$, and since $E(a,\zeta)$ and $E^\star(a,\zeta)$ are linearly independent, the transmission and reflection coefficients of the wave energy flux finally read, by identification,
\algn{
T^2=\left\vert \frac{t}{I}\right\vert^2= \frac{1}{1+e^{2 \pi a}} ~~~~~~\mbox{and}~~~~~~R^2=\left\vert \frac{r}{I} \right\vert ^2=\frac{e^{2\pi a}}{1+e^{2\pi a}} \; .
}
\begin{figure}
   \centering
 \includegraphics[width=\hsize, trim = 0.8cm 0cm 0.8cm 1cm,clip]{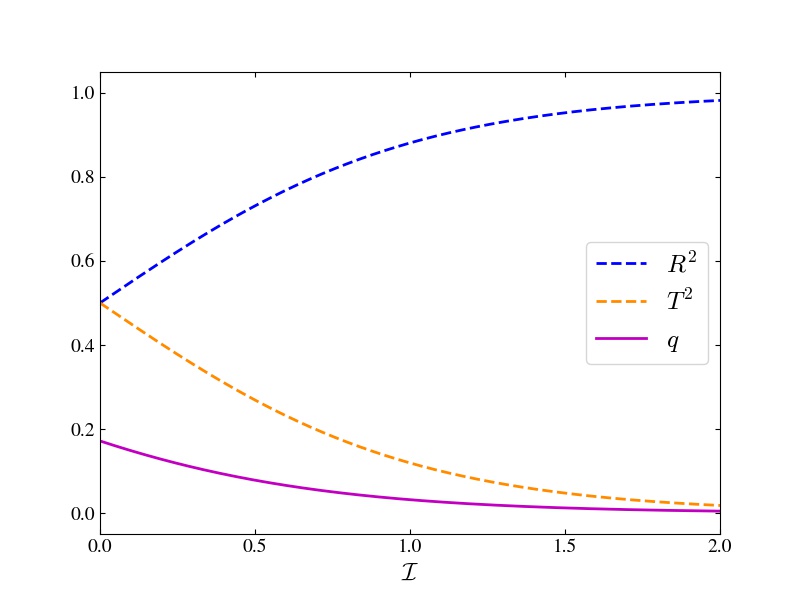}
      \caption{Squared wave reflection and transmission coefficients as a function of the wavenumber integral in the case of a parabolic barrier (blue and yellow dashed lines, respectively). The corresponding coupling factor computed with \protect\eq{def q} is also plotted in solid line. }
	  \label{q vs a}
\end{figure}
These expressions satisfy the energy conservation since \smash{$R ^2+ T ^2=1$.}
Using the wavenumber expression in \eq{Psi eq}, we note that the wavenumber integral over the evanescent region, $\mathcal{I}$, is related to the parameter $a$ by
\algn{\mathcal{I}=\int_{r_{\rm ev}-\Delta r/2}^{r_{\rm ev}+\Delta r/2} \vert k_r\vert\dd r=\frac{\pi}{4} \vert k_r(r_{\rm ev}) \vert \Delta r=\pi a \; , \label{I pi a}}
so that the transmission coefficient can be written as a function of $\mathcal{I}$ only, that is
\algn{T ^2 = \frac{1}{1+e^{2 \mathcal{I}}}\; . \label{T parabolic b}}
The wave coefficients and the associated coupling factor $q$ computed with \eq{def q} are displayed in \figurename{}~\ref{q vs a} as a function of $\mathcal{I}$.
As seen in \figurename{}~\ref{q vs a}, the smaller $\mathcal{I}$, the larger both $T^2$ and $q$ in the ideal case of a parabolic evanescent zone. The result is discussed in more detail within the framework of mixed mode coupling in \sectionname{}~\ref{toy model}.

%_________________________________

\section{Gradient-related term $\mathcal{G}_0$ in the formalism of \cite{Takata2016a}}
%_________________________________
\label{gradient term}
%------------------------------------------------------

%------------------------------------------------------
\subsection{General expression}
%-----------------------------------------------------
\label{G_0 general}

In the limiting case of a very thin evanescent zone, the asymptotic analyses of \cite{Takata2016a} resulted in
\algn{
\mathcal{G}_0&=\frac{\pi}{\kappa(x_0)}\left( \deriv{\ln c}{x}\right)_{x_0}^2 \; .
\label{def G_0}
}
In the latter, we have defined $x=\ln r$, \smash{$x_0=(x_2+x_1)/2$}, and \smash{$x_{\{1,2\}}=\ln r_{\{1,2\}}$}, where $r_1$ and $r_2$ are the turning points for which $\mathcal{N}(r_1)=\sigma$ and \smash{$\mathcal{S}_1(r_2)=\sigma$,} respectively. The other quantities are given by

\algn{
\kappa&=\sqrt{ST}\label{kappa}\\
c&=\left(\frac{S}{T}\right)^{1/4}\label{c takata}\\
S&=\frac{1}{(x_2-x)}\frac{2J}{F^2}\left(1-\frac{\sigma^2}{\mathcal{S}_1^2} \right)>0\label{S takata}\\
T&=\frac{J F^2}{(x-x_1)}\left(1-\frac{\mathcal{N}^2}{\sigma^2} \right) >0\label{T takata}\\
F&=\exp \left[ \frac{1}{2} \int^x\left(\mathcal{V}^\star-\mathcal{A}^\star  -J\right) {\rm d}x\right] \label{F takata}\; ,
}
where $J$ is defined in \eq{def J},
\algn{
\mathcal{V}^\star=-\frac{1}{\Gamma_1J} \deriv{\ln p}{x}~~~~\mbox{and}~~~~\mathcal{A}^\star=\frac{1}{J}\left(\frac{1}{\Gamma_1}\deriv{\ln p}{x}-\deriv{\ln \rho}{x} \right) \; .
\label{def A and V}
}
To go further, we also note that according to \eq{Lamb}{BVaisala}, $\mathcal{V}^\star$ and $\mathcal{A}^\star$ are respectively related to the (modified) Lamb and Brunt-Väisälä frequencies through
\algn{
\mathcal{V}^\star=\frac{1}{\Gamma_1J } \frac{ \rho g r}{p} =\frac{2 J}{\mathcal{S}_1^2}\frac{ g}{r}~~~~\mbox{and}~~~~\mathcal{A}^\star=J  \mathcal{N}^2\frac{ r}{g}  \; , \label{SVNA}
}
where the hydrostatic equilibrium was used in the first equality.

%-----------------------------------------------------
\subsection{Expression in a simple model}
%-----------------------------------------------------
\label{G_0 model}
In this section, we compute $\mathcal{G}_0$ for dipolar modes assuming that the modified Brunt-Väisälä and Lamb frequencies vary as power laws of radius with the same exponent, $\beta$. Following the assumptions in \tablename{}~\ref{table model}, we focus only on Type-a evanescent zones. To do so, we follow all the successive derivation steps described in \appendixname{}~A.3.1~and~A.3.2 of \cite{Takata2016a}. However, instead of considering that the thickness of the evanescent zone is null, we explicitly take its effect into account. The demonstration is adapted with our notation, that is, by replacing $\mu$ by $ 2 \beta$, and $s_0$ by $  -(\ln \alpha )/ 2\beta$, with $\alpha=(r_1/r_2)^{\beta}$.

First, using the profiles of $\mathcal{N}$ and $\mathcal{S}_1$ in \eq{power law N and S}, the resulting relation $J=2 \beta/3={\rm const.}$ (see \appendixname{}~\ref{equivalence}), and the fact that $r(x_0)=\sqrt{r_1 r_2}$, we obtain from \eq{kappa}{S takata}{T takata}
\algn{
\kappa(x_0)=\frac{4\sqrt{2}}{3}~\beta^2 \frac{(\alpha-1)}{\ln \alpha} \; .
\label{kappa b}
}
In a similar way, using \eqss{c takata}{F takata} and $J=2 \beta/3$, the derivative of $c$ gives
\algn{
\left( \deriv{\ln c}{x}\right)_{x_0} = -\beta \left(\frac{1}{\ln \alpha} +\frac{\alpha}{1-\alpha} \right) -\frac{\mathcal{V}^\star(x_0)-\mathcal{A}^\star(x_0)}{2}+\frac{ \beta}{3}\; .
\label{dc_ds b}
}
For the computation of $\mathcal{V}^\star(x_0)$ and $\mathcal{A}^\star(x_0)$, we follow \cite{Takata2016a} and use the assumptions (1) $\mathcal{V}^\star>\mathcal{A}^\star$ and \smash{(2) $\mathcal{V}^\star+\mathcal{A}^\star =3$.} For sake of clarity, we justify both assumptions.
First, the hypothesis (1) results from the fact that the equation is polytropic in a good approximation in the radiative zones of evolved stars so that the pressure and the density must satisfy $p\propto \rho ^\gamma$, with $\gamma \sim 1 - 4/3$ (e.g., see \appendixname{}~\ref{polytropic approx}). From \eq{def A and V}, we can therefore write \smash{$\mathcal{A}^\star \sim  \mathcal{V}^\star (\Gamma_1/\gamma-1)$}. For a totally ionized monatomic ideal gas, $\Gamma_1=5/3$ and we retrieve~(1).
Second, the hypothesis (2) comes from the assumption about the power-law profiles of $\mathcal{N}$ and $\mathcal{S}_1$. Indeed, such an hypothesis implies that $J$ must be considered as uniform, so that according to \eq{def A and V}{B=f(J)}, we retrieve~(2).
Noting also from \eq{power law N and S},~(\ref{SVNA}) and the relation $J=2 \beta/3$ that
\algn{
\left(\mathcal{V}^\star \mathcal{A}^\star \right)_{x_0} = \left(\frac{2 J^2 \mathcal{N}^2}{\mathcal{S}_1^2}\right)_{x_0}=\frac{8}{9}~  \beta^2 \alpha^2 \; ,
}
the use of the hypotheses (1) and (2) enables us to obtain $\mathcal{A}^\star (x_0)$ by solving the quadratic equation
\algn{\mathcal{A}^\star (x_0)^2-3\mathcal{A}^\star (x_0)+\frac{8}{9}~  \beta^2 \alpha^2=0\; ,}
which gives
\algn{
\mathcal{A}^\star (x_0) = \frac{3}{2} - \beta \sqrt{\frac{9}{4 \beta^2}-\frac{8 \alpha^2}{9}} \; .
\label{last soon}
}
Finally, considering \eq{last soon} and $\mathcal{V}^\star+\mathcal{A}^\star =3$ in \eq{dc_ds b}, and considering \eq{kappa b}, we find that the gradient-term $\mathcal{G}_0$ given in \eq{def G_0} reads in the simple considered model 
\algn{
\mathcal{G}_0=\frac{3\pi}{4\sqrt{2} } \frac{\ln \alpha}{(\alpha-1)}  \left( \frac{1}{\ln \alpha} +\frac{\alpha}{1-\alpha}-\frac{1}{3}+\sqrt{ \frac{9}{4 \beta^2}-\frac{8\alpha^2}{9}}~\right)^2  \; .
\label{G_0 alpha} 
}

In the limit of $\Delta r\rightarrow 0$, that is, around $\alpha =1$, it is possible to check using a second-order Taylor expansion that \eq{G_0 alpha} tends  to
\algn{
\mathcal{G}_0 (\alpha \rightarrow 1)=\frac{\pi}{12\sqrt{2}} \left( \frac{5}{2}-\sqrt{\frac{81}{4\beta^2}-8 }~\right)^{2} \; ,
\label{G_0 M17}
}
which corresponds to the expression found by \cite{Takata2016a}. Besides, the extrapolation of $\mathcal{G}_0$ outside of its derivation domain, that is, in the limit of $\alpha\rightarrow 0$, is equal according to \eq{G_0 alpha} to
\algn{
\mathcal{G}_0 (\alpha \rightarrow 0)\approx -\frac{3\pi}{4\sqrt{2} } \left( \frac{3}{2 \beta} -\frac{1}{3}\right)^2  \ln \alpha  \; .
\label{G_0 approx}
}

\end{document}